\documentclass[11pt,a4paper]{article} 
\usepackage{heppub} 

\usepackage{amsmath,cancel,mathtools, braket, slashed, placeins, multirow, colortbl, makecell, marginnote}
\usepackage[normalem]{ulem}
\usepackage{subcaption}
\usepackage[T1]{fontenc}
\usepackage[utf8]{inputenc}
\usepackage[usenames,dvipsnames]{xcolor}

\DeclareRobustCommand{\Res}{\text{Res}}

\DeclareRobustCommand{\bub}{\text{bub}}
\DeclareRobustCommand{\Ebar}{\ensuremath{\bar{E}}}
\DeclareRobustCommand{\lbar}{\ensuremath{\bar{l}}}
\DeclareRobustCommand{\pbar}{\ensuremath{\bar{p}}} 
\DeclareRobustCommand{\qbar}{\ensuremath{\bar{q}}}

\DeclareRobustCommand{\cM}{\ensuremath{\mathcal{M}}}
\DeclareRobustCommand{\cO}{\ensuremath{\mathcal{O}}}
\newcommand{\MSbar}{{\overline{\mathrm{MS}}}}

\newcommand{\nn}{\nonumber}

\title{Renormalons in the energy-energy correlator}

\author[a]{Stella T.~Schindler,}
\author[a]{Iain W.~Stewart,} 
\author[a]{and Zhiquan Sun} 

\affiliation[a]{Center for Theoretical Physics, Massachusetts Institute of Technology\\
	77 Massachusetts Ave., Cambridge, MA 02139, USA}

\emailAdd{stellas@mit.edu}
\emailAdd{iains@mit.edu}
\emailAdd{zqsun@mit.edu}

\abstract{
The energy-energy correlator (EEC) is an observable of wide interest for collider physics and Standard Model measurements, due to both its simple theoretical description in terms of the energy-momentum tensor and its novel features for experimental studies. 
Significant progress has been made in both applications and higher-order perturbative predictions for the EEC.
Here, we analyze the nature of the asymptotic perturbative series for the EEC by determining its analytic form in Borel space under the bubble-sum approximation.  This result provides information on the leading and subleading nonperturbative power corrections through renormalon poles. 
We improve the perturbative convergence of the $\overline{\mathrm{MS}}$ series for the EEC by removing its leading renormalon using an R scheme, which is independent of the bubble-sum approximation. Using the leading R-scheme power correction determined by fits to thrust, we find good agreement with EEC OPAL data already at ${\mathcal O}(\alpha_s^2)$. 
}

\date{May 30, 2023}

\preprint{\vbox{
		\hbox{MIT-CTP 5499}
}}

\arxivnumber{}

\begin{document}
	\allowdisplaybreaks
	\maketitle

\section{Introduction}

A central challenge of high-energy physics is to develop observables that are easily accessible experimentally, 
have a clean and clear connection to dynamics of interest in the underlying physical theory, 
and can be computed theoretically at high precision. 
An important observable for quantum chromodynamics that currently satisfies the first two criteria is the energy-energy correlator (EEC),
an infrared and collinear (IRC) safe event shape that characterizes the angular distribution of particles produced in $e^+e^-$ collisions \cite{PhysRevLett.41.1585, PhysRevD.19.2018},
\begin{align}
  \frac{d \Sigma}{d \cos\chi} = \sum_{i,j} \int d \sigma \frac{E_i E_j}{Q^2} \delta(\cos\chi-\cos\theta_{ij}) \,.
\end{align} 
The EEC has many useful generalizations, including the transverse EEC (TEEC) for hadron-hadron collisions \cite{Ali:1984yp}, multi-point energy correlators \cite{Chen:2019bpb}, as well as EEC and TEEC observables for electron-hadron colliders~\cite{Li:2021txc, Li:2020bub}; it is also related to energy correlation functions for jets \cite{Chen:2020vvp,Komiske:2022enw,Lee:2022ige}. 
Nonperturbative contributions to energy correlators are expected to be small, making them compelling candidates for studying strong interactions \cite{Neill:2022lqx}, whether through extracting the QCD coupling constant \cite{Burrows:1995vt, Kardos:2018kqj}, investigating transverse momentum distributions \cite{Li:2020bub,Li:2021txc}, or probing factorization violation \cite{Gao:2019ojf}.
These observables have been measured at LEP and the LHC at CERN \cite{OPAL:1990reb,ALEPH:1990vew,L3:1991qlf, ATLAS:2015yaa,ATLAS:2017qir,ATLAS:2020mee}, SLD at SLAC \cite{SLD:1994yoe}, 
and are a target of the planned Electron-Ion Collider at Brookhaven \cite{AbdulKhalek:2021gbh}.
To interpret experimental data, it is important to have corresponding high-precision theoretical predictions.

The EEC was first studied at leading order in the 1970s \cite{PhysRevD.19.2018},
instigating a decade of work producing numerical predictions at next-to-leading order (NLO) 
\cite{Schneider:1983iu, Falck:1988gb, Glover:1994vz, Kramer:1996qr, Ali:1982ub, Ali:1984gzn, Richards:1982te, Richards:1983sr, Catani:1996jh}.
It took two further decades to achieve the first numerical results at next-to-next-to-leading order (NNLO)  \cite{DelDuca:2016csb, Tulipant:2017ybb} and analytic results at NLO \cite{Dixon:2018qgp}.
Higher-order perturbative results for the EEC exist only for certain cases: 
in the so-called collinear limit at next-to-next-to-leading logarithms (NNLL) \cite{Dixon:2019uzg, Korchemsky:2019nzm, Chen:2020uvt}, 
in the back-to-back limit at N$^3$LO, N$^3$LL$'$ and N$^4$LL~\cite{Kodaira:1981nh, Kodaira:1982az, deFlorian:2004mp, Tulipant:2017ybb, Moult:2018jzp, Ebert:2020sfi,Duhr:2022yyp}, 
as well as for $\mathcal{N}=4$ supersymmetric Yang-Mills theory both at leading power at NNLO and at subleading power only in the back-to-back limit \cite{Belitsky:2013xxa, Belitsky:2013bja, Belitsky:2013ofa, Henn:2019gkr, Moult:2019vou}. 

There has been much less recent work on nonperturbative power corrections for the EEC.  
\Refcite{Nason:1995np} pointed out that the EEC has $\Lambda_{\rm QCD}/Q$ power corrections for all values of the angular variable $\chi$, in constrast to event shapes for which such power corrections appear in the dijet limit. Ref.~\cite{Nason:1995np} also carried out a renormalon calculation for a weighted integral over the EEC, finding a result in agreement with the presence of $1/Q$ power corrections.  \Refcite{Korchemsky:1999kt} initiated the use of operator methods to study EEC power corrections, and argued that the leading corrections take the form 
\begin{align} \label{eq:KSpower}
   \frac{1}{\sigma_0} \frac{d \Sigma}{d \cos\chi} 
  =  \frac{1}{\sigma_0} \frac{d \hat \Sigma}{d \cos\chi} 
   + \frac{2}{\sin^3\!\chi} \frac{\bar\Omega_1}{Q} 
  \,,
\end{align}
with $\bar\Omega_1\sim \Lambda_{\rm QCD}$.  Here $\bar\Omega_1=\lambda_1/2$ in the notation of \refcite{Korchemsky:1999kt,Belitsky:2001gf}. Note that relative to~\cite{{Korchemsky:1999kt,Belitsky:2001gf}}, we multiply by $2\pi$ for the azimuthal angular integral and include an additional combinatoric factor of $2$, since either of the energies $E_i$ or $E_j$ can become nonperturbative. 
\Refcite{Dokshitzer:1999py} reached this same conclusion in the dispersive model for power corrections with the same factors of $2$ shown in \eq{KSpower},  and also considered perturbative corrections to the power correction coefficient. The universality of the power correction is not violated by gluon splitting~\cite{Dokshitzer:1998pt}, an effect known as the Milan factor~\cite{Dokshitzer:1997iz}. The earliest prediction for the $1/\sin^3\chi$ behavior of EEC power corrections was in the fragmentation model of \refcite{PhysRevD.19.2018}.
Our $\bar\Omega_1$ notation follows \refcite{Abbate:2010xh}, where a value of $\bar\Omega_1$ for thrust was determined by a fit to data. 
Using massless kinematics, the parameter $\bar\Omega_1$ appearing for the EEC is the same as the one appearing in the dijet limit of thrust and other $e^+e^-$ event shapes~\cite{Korchemsky:1999kt,Belitsky:2001gf,Lee:2006nr}; however, this universality can be spoiled by ${\cal O}(1)$ hadron mass corrections, which depend on the method used to reconstruct particle energies and momenta~\cite{Salam:2001bd, Mateu:2012nk}.

Interestingly, to our knowledge, a direct bubble-sum renormalon calculation of the $\chi$-dependence of the $1/Q$ power correction to the EEC in Borel space has not yet been carried out, so we do so here. 
Our expectation is that the leading Borel-space renormalon will agree with the $1/\sin^3\!\chi$ in \eq{KSpower}; however, the renormalon analysis method relies on completely different approximations than the operator-based approach, and thus provides an independent nontrivial confirmation. 
The Borel-space result also gives access to higher-order renormalons. We find that in the bubble-sum approximation, there is no renormalon corresponding to an $\cO(1/Q^2)$ power correction. 
Additionally, our calculations provide input to assess the convergence of  the $\MSbar$-scheme perturbative expansion for the EEC, which is known to be asymptotic. At ${\cal O}(\alpha_s^2)$, we observe that the large-$\beta_0$ approximation provides a reasonable approximation to the full result.
From fig.~1 of~\refcite{Tulipant:2017ybb} at ${\cal O}(\alpha_s^3)$, we note that the EEC seems to exhibit slow perturbative convergence for some values of $\chi$.  This is also true for the $z\to 0$ fixed-order and resummed results in \refcite{Dixon:2019uzg}.

We present a scheme change from $\MSbar$ to an R scheme~\cite{Hoang:2009yr,Abbate:2010xh,Bachu:2020nqn}, which removes the leading renormalon from both the perturbative EEC series and its leading power correction. This scheme change yields improved convergence for the perturbative EEC series. Using the universality between the thrust and EEC $\bar\Omega_1$ parameters, we show that our R scheme results are already consistent at ${\cal O}(\alpha_s^2)$ with EEC data from the OPAL experiment at LEP \cite{OPAL:1993pnw}. 

The outline of this paper is as follows. 
\Sec{borel} provides a brief overview of asymptotic series and the types of divergences that they may exhibit, including renormalons. 
In \sec{calculation}, we utilize the bubble chain formalism 
for probing renormalons from the infrared structure of final-state gluons to calculate a Borel space result for the EEC, and demonstrate that its leading renormalon is consistent with the $1/Q$ power correction in \eq{KSpower}, while a renormalon corresponding to the $1/Q^2$ power correction is absent.
In \sec{msr}, we show how to remove the $u=1/2$ renormalon from the $\MSbar$ perturbative series using an R scheme, and demonstrate that this improves both perturbative convergence and agreement with experimental data.
We make concluding remarks in \sec{conclusion}.

\subsection{Borel summation and renormalons}\label{sec:borel}

Perturbative expansions of observables in QFT are generally asymptotic \cite{Dyson:1952tj}:
even if their first few terms appear to converge, at some order of expansion, the coefficients begin to rapidly grow. 
Divergences should not be viewed as a fundamental sickness of a theory; 
rather, asymptotic series encode all-orders nonperturbative information \cite{Bender:1969si,Bender:1973rz, Beneke:1998ui,Argyres:2012ka, Dunne:2013ada}.
Nonetheless, extracting useful information from such a series is nontrivial; 
there are many different types of divergences, each of which must be handled in different ways.

Perhaps the most conspicuous source of divergence in QFT  is factorial growth in the number of Feynman diagrams at high orders, which may lead to corresponding growth of the series \cite{Hurst:1952zh, Bender:1976ni, Lipatov:1976ny, Zinn-Justin:1980oco}. 
The technique of Borel summation is often sufficient to overcome such growth. 
First, we take a Borel transform of an observable $f(\alpha_s) \to F(u)$, which for QCD is most often defined by:
\begin{align}\label{eq:borel}
	f(\alpha_s) &= \sum_{n=-1}^\infty  c_n\, \alpha_s^{n+1}  \quad \to \quad 
	F(u) =  c_{-1}\, \delta(u) + \sum_{n=0}^\infty \frac{c_n}{n!}  \Big( \frac{4\pi}{\beta_0} \Big)^{n+1}\, u^n\,.
\end{align}
The factor of $4\pi/\beta_0$ provides a convenient normalization for the Borel variable $u$, where $\beta_0=11 C_A/3 - 2 n_f/3$ is the lowest order QCD $\beta$-function.
Next, we compute the new, more rapidly convergent sum $F(u)$. 
Finally, we take an inverse Borel transform to recover the initial observable $f(\alpha_s)$: 
\begin{align}\label{eq:inverse-borel}
	f(\alpha_s) &= \int_0^\infty du \, \exp\Big[-u \frac{4\pi}{\beta_0\alpha_s(\mu)}\Big]\, F(u)  \,.
\end{align}
For more details of summation methods, see \refscite{Bender_Orszag_1999, Beneke:1998ui}.

Borel summation techniques are powerful but not sufficient for every series.
If an observable exhibits poles on the positive real axis in the complex-$u$ Borel plane \cite{Gross:1974jv, Lautrup:1977hs, tHooft:1977xjm},
the inverse Borel transform runs into a problem. 
In order to evaluate the integral in \eq{inverse-borel}, 
one must deform the integration contour above or below the real line to avoid the pole. 
Depending on where one chooses to deform the contour, 
the inverse Borel transform will yield different results.
This leads to an ambiguity in the value of the perturbative series,
the magnitude of which is given by the residue of the pole.
Such poles are called infrared (IR) renormalons, which we hereafter refer to as simply renormalons (poles on the negative real axis are UV renormalons, which are not discussed here).
Renormalons nearer to the origin create greater uncertainty in the value of a perturbative series, since the series coefficients grow as $\sim n!\, (2/p)^{n}$ for large $n$, given a pole at $u=p/2$. 

Generally, one wishes to evaluate QCD observables theoretically at a level of precision comparable to experiment. 
Luckily, renormalon ambiguities in observables are unphysical: they cancel out between the perturbative series and nonpeturbative matrix elements (or parameters of the field theory). 
We may view them as merely an artifact of not cleanly separating perturbative and nonperturbative contributions to an observable when carrying out a coupling expansion.%
\footnote{The path integral formalism provides further intuition for divergent series in QFT.
We may conceptualize perturbative effects as representing fluctuations about the vacuum, 
whereas nonperturbative effects arise from fluctuations about nontrivial saddle points. 
Some nonperturbative saddle points, such as instanton-anti-instanton pairs, are Borel summable \cite{ Bogomolny:1977ty, Behtash:2018voa}.
On the other hand, renormalons are connected to the more nuanced bions, or fractional instanton-anti-instanton pairs
\cite{Argyres:2012ka, Dunne:2012zk, Dunne:2013ada, Cherman:2013yfa, Basar:2013eka, Dunne:2014bca, Cherman:2014ofa, Fujimori:2018kqp,Unsal:2021cch}.
Developments from the rich and evolving field of hyper-asymptotics, resurgence theory, and trans-series 
\cite{Ecalle:1981:FRTa,boyd_devils_1999,Dorigoni:2014hea, Aniceto:2018bis} 
promise to provide further insight into nonperturbative effects in QFT as well as mathematical tools for handling asymptotic expansions unambiguously.

To construct a trans-series, one sums up a traditional perturbative series with a nonperturbative contribution, expressed as a complicated sum involving powers of exponential and logarithmic functions of the coupling. 
If one examines only a subset of terms in the trans-series, an imaginary ambiguity like a renormalon may arise. However, one expects~\cite{Ecalle:1981:FRTa} that a set of the so-called resurgence relations connecting the perturbative and nonperturbative sectors of the trans-series will cancel all ambiguities in the overall observable.
This appears to be related to well-known phenomena in QCD where renormalon ambiguities are removed by using alternate renormalization schemes to define the perturbative and nonperturbative terms (see~\cite{Beneke:1998ui,Manohar:2000dt} for reviews), which we explore later in this paper with an R scheme.
}
 
Renormalons originate in the IR region of momentum integrals in loop diagrams \cite{Beneke:1998ui}, and their behavior depends crucially on our choice of renormalization scheme; they appear in many schemes, such as the popular $\MSbar$ scheme.
We may improve the convergence of a perturbative expansion by modifying our renormalization scheme choice for matrix elements (or parameters) in a manner that systematically removes these instabilities. This modification improves both perturbative behavior and the stability of the matrix element (or parameter) extraction.  
Examples of renormalons in QCD include
the heavy-quark mass \cite{Bigi:1994ng, Beneke:1994sw, Bigi:1997ty, Pineda:2001ia, Hoang:2008yj}, 
the $B^*$-$B$ mass splitting governed by the parameter $\lambda_2$~\cite{Grozin:1997sa,Hoang:2009yr}, 
and hadronization parameters describing jet cross-sections~\cite{Mueller:1984vh,Beneke:1994sw,Dasgupta:2003iq,Fleming:2008po,Abbate:2010xh,Hoang:2015gj,Gracia:2021nut}.
In this paper, we obtain a Borel-space result for EEC renormalons, and then apply the method of \refcite{Hoang:2009yr} to improve stability for the EEC.

\section{EEC renormalons}\label{sec:calculation}
The EEC describes how the energies of final-state hadrons in an $e^+e^-$ collision are correlated, as a function of their angle $\chi$ relative to one another:
\begin{align}\label{eq:eec-definition}
	\frac{d\Sigma}{d\cos\chi} = \sum_{i,j}\int d\sigma_{e^+e^-\to ijX} \frac{E_i E_j}{Q^2}\delta(\cos\chi-\cos\theta_{ij}).
\end{align}
The right-hand side of \eq{eec-definition} is a weighted sum over differential cross-sections for all possible inclusive processes $e^+e^-\to ijX$
in which particles $i$ and $j$ are detected, with $X$ representing the remaining (arbitrary) final states. 
We take the total invariant mass of the incoming $e^+e^-$ to be $Q^2$ and work in the center-of-mass frame,
where the measured final-state particles carry energy $E_i$ and $E_j$ and are separated by angle $\theta_{ij}$.

\begin{figure}[t]
	\begin{center}
		\begin{subfigure}{2in}\includegraphics[width = 2 in]{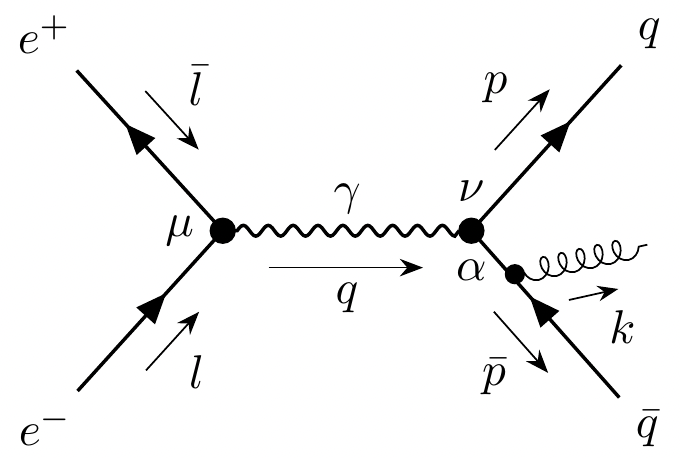}
		\caption{}\label{fig:eec-lo-a}\end{subfigure}
		\qquad\qquad
		\begin{subfigure}{2in}\includegraphics[width = 2 in]{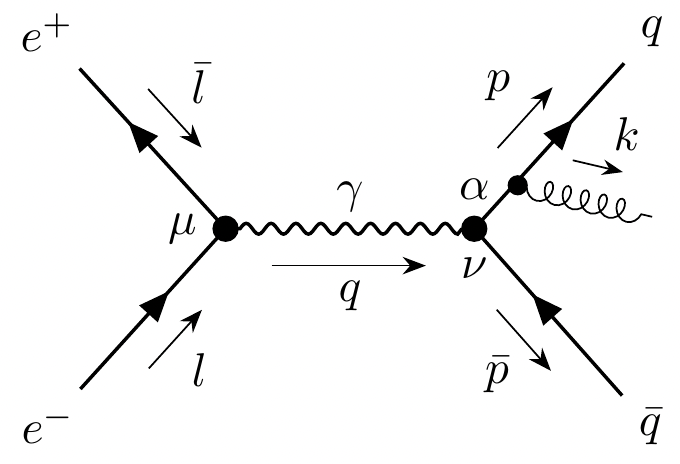}
		\caption{}\label{fig:eec-lo-b}\end{subfigure}
	\end{center}
\vspace{-0.3cm}
	\caption{Leading-order diagrams for the energy-energy correlator in \eq{eec-definition}.}
  \label{fig:eec-at-lo}
\end{figure}

\subsection{Leading order}

We begin by briefly reviewing the calculation of the EEC at leading order,
as our renormalon studies directly build off of this result. 
At LO, the EEC is defined in terms of $e^+e^-\to q\qbar g$ Feynman diagrams shown in \fig{eec-at-lo}, 
and \eq{eec-definition} becomes 
\begin{align}\label{eq:eec-at-lo}
	\frac{d\hat\Sigma}{dz} 
	= \int\!\! \frac{d^3p}{2E}\frac{d^3\pbar}{2\Ebar} \frac{d^3k}{2E_g} 
	\frac{\delta^4(q-p-\pbar-k)}{(2\pi)^{5}}  \sum_{i,j=q,\qbar, g} 
	\frac{E_i E_j}{Q^4}
	\delta(z-z_{ij})
    \alpha_s(\mu)\, \langle|\cM_0|^2\rangle,
\end{align}
where we write the angular variable as $\cos\chi = 1-2z$ and $\cos\theta_{ij} = 1-2z_{ij}$.
Here, $p$, $\pbar$, and $k$ are the four-momenta of a final-state quark, anti-quark, and gluon, respectively, 
with corresponding energies $E$, $\Ebar$, and $E_g$. 
The four-momentum of the virtual photon is $q$, with $Q^2 = q^2$.
The initial-state leptons have four-momenta $l$ and $\lbar$, and treating the leptons as unpolarized, we can average over both spins and leptonic angles relative to a given coordinate system.
The spin-averaged LO matrix element $\langle |\cM_0|^2\rangle$ from summing \figs{eec-lo-a}{eec-lo-b} then evaluates to 
\begin{align}\label{eq:no-bubble-squared}
	\langle | \cM_0|^2\rangle 
&= \sigma_0\, \pi^2 2^{9}  C_F 
  \frac{E^2+ \Ebar^2}{(Q-2E)(Q-2\Ebar)} \,.
\end{align}
Here, the tree-level cross section is
\begin{align}
  \sigma_0 = \frac{4\pi\alpha^2}{Q^2} \sum_q e_q^2 \,,
\end{align}
where $e_q$ is the charge of a quark of flavor $q$. 
Plugging \eq{no-bubble-squared} into \eq{eec-at-lo} and carrying out the phase space integrals, one finds the result for the EEC at $\cO(\alpha_s)$:
\begin{align}\label{eq:lo-value}
	\frac{1}{\sigma_0}	\frac{d\hat\Sigma_{\rm LO}}{dz} 
= \frac{\alpha_s C_F}{4 \pi}\frac{3-2z}{z^5(1-z)} \Big[3z(2-3z) + 2(3-6z+2z^2)\log(1-z) \Big] \,.
\end{align}
Note that this cross-section scales as $1/z$ as $z\to 0$ and as $1/(1-z)$ as $z\to 1$ (modulo logarithms), and that this is the generic scaling expected for perturbative contributions to the EEC in these limits.

\subsection{Bubble diagram calculation}\label{sec:bubbles}

\begin{figure}
	\begin{center}
		\includegraphics[width = 3.5 in]{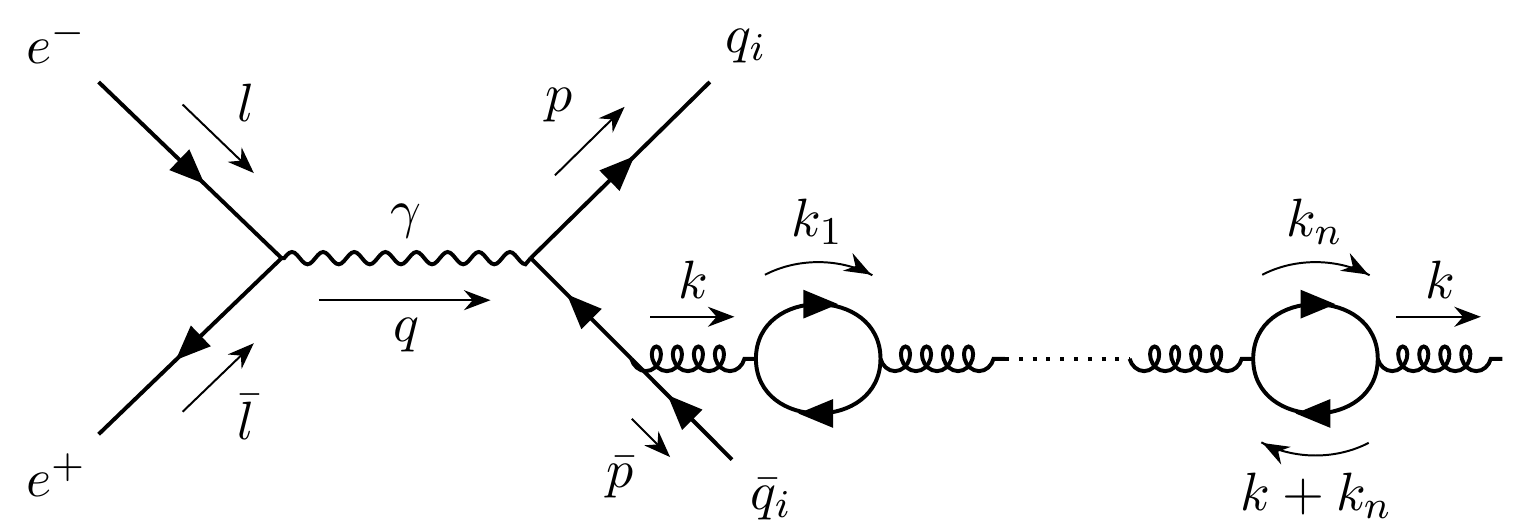}
	\end{center}
	\caption{
		EEC bubble diagram, formed by replacing the external gluon in \fig{eec-lo-a} with a chain of $n$ fermion bubbles. 
		Evaluating \eq{eec-bubs} over the sum of all bubble diagrams allows us to probe for renormalon divergences in the EEC. 
	}\label{fig:bubbles}
\end{figure}

Computing all possible radiative corrections to an observable and analyzing them for singularities is prohibitively involved.
In QCD, it is conventional to instead examine the smaller set of so-called bubble diagrams to probe the existence of a renormalon. 
We define an $n$-bubble diagram by replacing the outgoing gluon in \fig{eec-lo-a} and \fig{eec-lo-b} by a chain of $n$ fermion loops, shown in detail in \fig{bubbles}.
We must evaluate \eq{eec-at-lo} with $\cM_0$ replaced by the sum $\cM_\bub$ of all possible $n$-bubble diagrams, from $n=0$ to $\infty$:
\begin{align}\label{eq:eec-bubs}
	\frac{d\hat\Sigma_\bub}{dz}
	=\int\!\! \frac{d^3p}{2E}\frac{d^3\pbar}{2\Ebar} \frac{d^3k}{2E_g} 
	\frac{ \delta^4(q-\sum p_i)}{(2\pi)^{5}} \sum_{i,j}
	\frac{E_i E_j}{Q^4}
	\delta(z-z_{ij})\alpha_s(\mu)\, \langle|\cM_\bub|^2\rangle,
\end{align}
where
\begin{align}
	i \sqrt{\alpha_s}\cM_\bub = \sum_{n=0}^\infty \left[\,\,\raisebox{-0.45\height}
	{\includegraphics[width = 2 in]{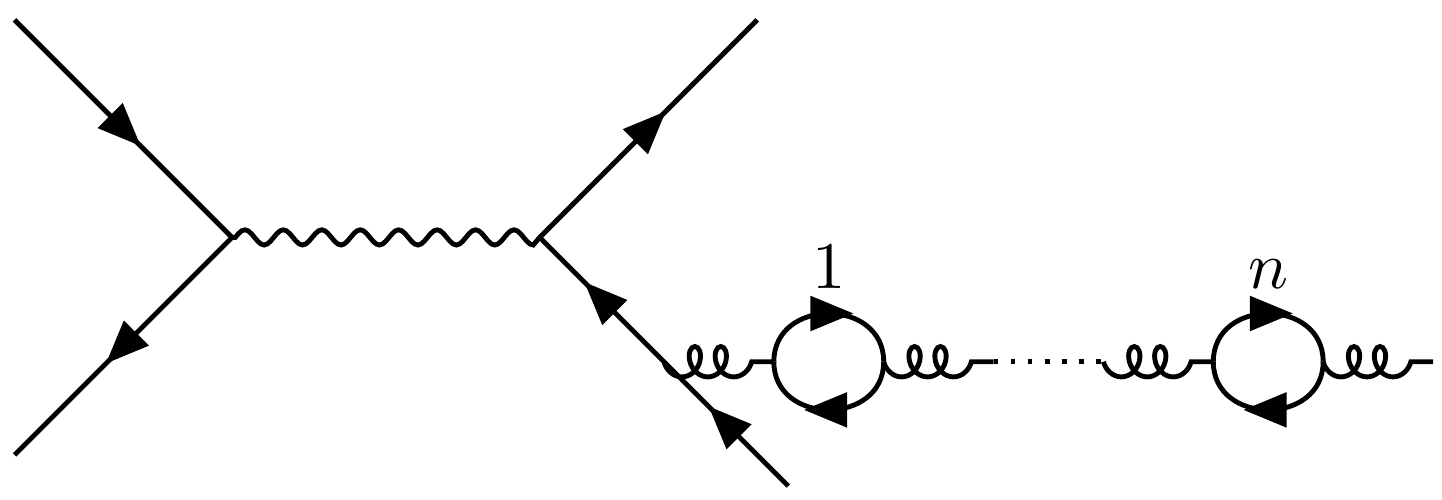}
	}
	 + \raisebox{-0.45\height}
	{\includegraphics[width = 2 in]{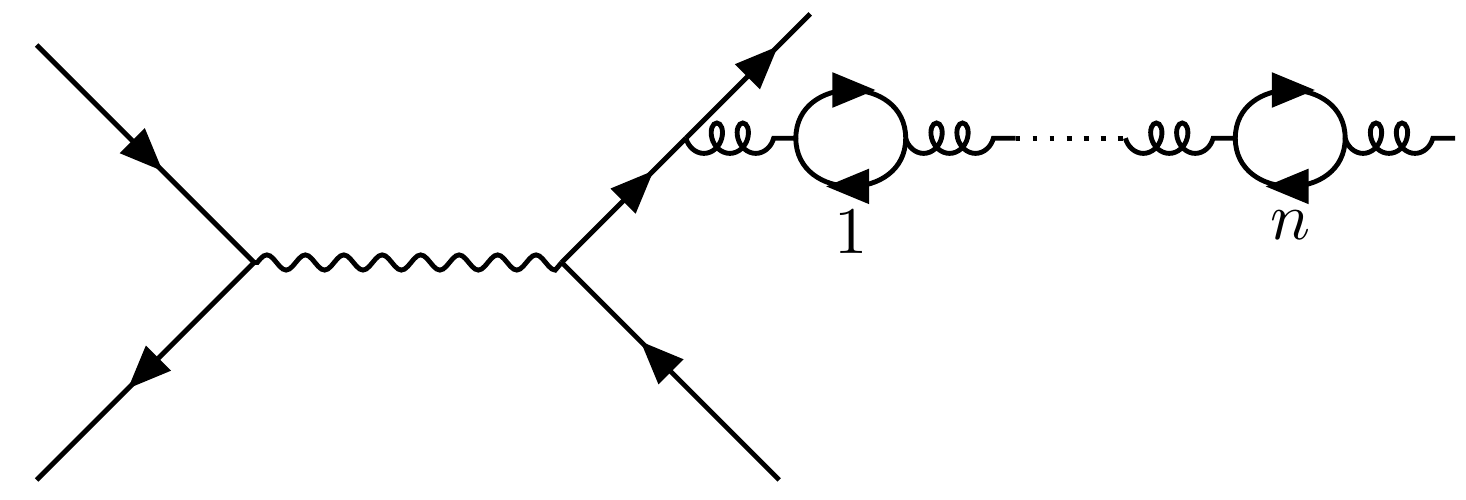}
	}\,\,\right].
\end{align}
Bubble diagrams comprise a gauge-invariant subset of all possible Feynman diagrams, which for $n_f$ light quark flavors are proportional to $n_f^k$, with the highest possible power of $k$ at each order in $\alpha_s$.
Thus, they serve as a convenient probe for the existence and severity of renormalon ambiguities \cite{Beneke:1998ui}. 
Though the bubble diagram procedure is by nature imprecise,
it nevertheless provides us sufficient information to both analyze and improve perturbative convergence, as well as better understand subleading power corrections.

It will be convenient to organize results at higher orders in perturbation theory in terms of color coefficients that involve $\{\beta_0,\,C_F,\,C_A,\,...\}$ rather than $\{n_f,\,C_F, \,C_A,\,...\}$. Here, $\beta_0 = 11C_A/3-2n_f/3$, and the ellipses denote terms beyond quadratic Casimirs. The basis with $\beta_0$ is more convenient here because terms with the maximum number of $\beta_0$ factors are known to numerically dominate perturbative series in many cases \cite{Brodsky:1982gc}, while terms with $C_F$ or $C_A$ in place of a $\beta_0$ are numerically subleading. The renormalon bubble-chain calculation predicts precisely these leading-$\beta_0$ terms.
Using the analytic results  at ${\cal O}(\alpha_s^2)$ from  \refcite{Dixon:2018qgp}, one can confirm that for the EEC the $\beta_0$ terms numerically dominate, as we show in \fig{contributions}.

\subsubsection{Modified gluon propagator probe}

Our impetus to evaluate \eq{eec-bubs} is that a bubble-sum-modified gluon propagator probes the infrared structure of the perturbative series, and thus can reveal renormalons.
To make the gluon propagator explicit, we rewrite the gluon phase-space integral in \eq{eec-bubs} as
\begin{align}\label{eq:gluon-appears}
	\int \frac{d^3k}{2E_g}  
  = \: 2\int \frac{d^4k}{2\pi}\, \Theta(k^0) \, \text{Im} \left[ \frac{1}{-k^2-i0}\right] \,.
\end{align}
For expediency, we define $\Pi_g(k)$ as the gluon propagator times $g^2$, i.e. $ - (4\pi\alpha_s) i S^{\alpha\beta} /(k^2 + i0) = -4\pi i S^{\alpha\beta} \Pi_g(k)$. 
It is convenient to work in Lorentz gauge, so that the same transverse Lorentz index structure is present at tree level and in the presence of quark bubbles, in which case $S^{\alpha\beta} = (g^{\alpha\beta}-{k^\alpha k^\beta}/{k^2})\delta^{ab}$.
The effect of inserting bubbles into \eq{eec-bubs} is
\begin{align}\label{eq:substitution}
	\langle |M_\bub|\rangle^2 \Pi_g (k) = \langle |M_0|\rangle^2 \Pi_\bub(k) 
\,, \end{align}
where $\Pi_\bub$ is the bubble-modified gluon propagator defined as
\begin{align}\label{eq:n-bubbles}
	- 4\pi i\, S^{\alpha\beta}\,\Pi_{\bub}(k) = \sum_{n=0}^\infty\raisebox{-0.4\height}
	{\includegraphics[width = 1.4 in]{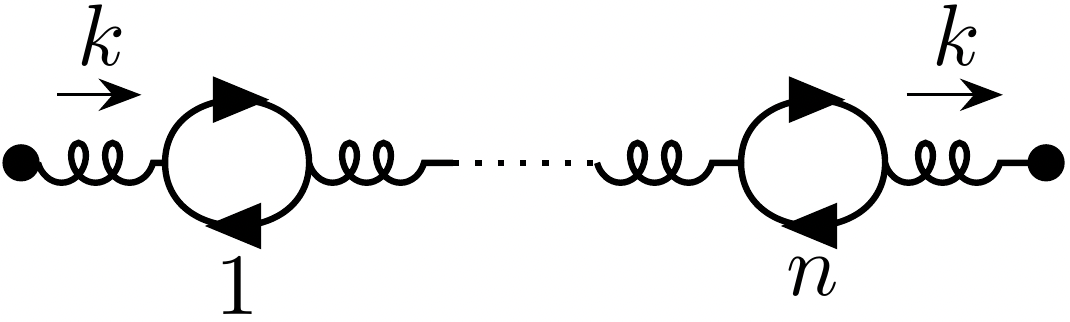}
	} 
\,, \end{align}
where the dots on the ends give a factor of $g^2 = 4\pi\alpha_s$.
Substituting \eq{substitution} into \eq{eec-bubs},
\begin{align}\label{eq:eec-bubs-1}
	\frac{d\hat\Sigma_{\bub}}{dz}=2\!\!\int\!\! \frac{d^3p}{2E}\frac{d^3\pbar}{2\Ebar}\frac{d^4k}{2\pi} \Theta(k^0) \, 
     \sum_{i,j}	\frac{E_i E_j}{Q^4}	\delta(z-z_{ij})
    \text{Im} \big[\!\!-\!\!\Pi_\bub\big]
	\frac{ \delta^4(q-\sum p_i)}{(2\pi)^{5}} \langle|\cM_0|^2\rangle.
\end{align}

We will make an approximation when evaluating \eq{eec-bubs-1}.  Diagrammatically, the result for Im$\,\Pi_\bub$ includes both terms with a cut gluon propagator and terms where a pair of quarks in a bubble are cut. While both of these cuts are included in our analysis, we allow the energy factors $E_i$ and $E_j$ to select only from the original $q$, $\bar q$, or a gluon with $k^0$, and never from the individual quarks in the cut quark bubble. 
This has the virtue of significantly simplifying the calculation while still allowing the resulting modified gluon propagator to act as a probe of infrared dynamics.

Let us evaluate the series for $\Pi_\bub$ in \eq{n-bubbles}.
The $n=0$ term is $-4\pi i S^{\alpha\beta}\Pi_g$. 
The $n=1$ renormalized result is well known:
\begin{align}
	\raisebox{-0.45\height}{\includegraphics[width = 1.7 in]{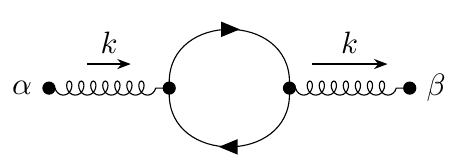} 
	} \,
	&= -\frac{2\alpha_s^{2} n_f}{3}\frac{(-i S^{\alpha\beta})}{k^2+i0} \ln \Big(\frac{\mu^2e^c}{-k^2-i0} \Big)
  \,.
\end{align}
Again, we have included couplings and gluon propagators at either end. 
We 
use the modified minimal subtraction ($\MSbar$) renormalization scheme with parameter $\mu$, 
and define $c = 5/3$. 
It is now straightforward to evaluate the bubble propagator sum in (\ref{eq:n-bubbles}):
\begin{align}\label{eq:new-propagator}
	\Pi_\bub(k) = \frac{1}{k^2+i0} \frac{4 \pi}{\beta_{0}} \sum_{n=0}^\infty
\Big(\frac{\alpha_s\beta_0}{4\pi}\Big)^{n+1} 
\bigg[\ln \Big(\frac{\mu^2e^c}{-k^2-i0} \Big)\bigg]^n
\,. \end{align}
Here, we moved into the color coefficient basis $\{\beta_0, C_A, C_F,\ldots\}$ by making the standard replacement $n_f\to -3\beta_0/2$.
Next, we move \eq{new-propagator} into Borel space using \eq{borel},  taking $\alpha_s \beta_0/(4\pi)$ to the Borel variable $u$:
\begin{align}\label{eq:new-propagator-borel}
	\Pi_\bub(u) = \frac{-1}{(-k^2-i0)^{1+u}} \frac{4 \pi}{\beta_{0}} (\mu^2e^c)^u,
\end{align}
We can now convert the bubble phase-space integral in \eq{eec-bubs-1} back from four dimensions into a three-dimensional phase-space integral. When doing so, it is important to remember that the squared amplitude $\langle |{\cal M}_0|^2\rangle$ and EEC weighting factor $E_i E_j$ are functions of energies, which we denote by ${\cal A}(k^0)$.  For any ${\cal A}(k^0)$ we can carry out the transverse momentum integral to 
obtain\footnote{If we were to ignore ${\cal A}(k^0)$, it would be natural to carry out the $k^0$ integral to find the modified phase-space integral in the last line of \eq{bubble-4d}. 
This choice corresponds to using the equation of motion $k^0=|\vec k|$ to simplify ${\cal A}(k^0)$ before (rather than after) carrying out the bubble sum in Borel space. 
Because \eq{bubble-4d} involves an integral over offshell momenta $k^2\ne 0$, these choices lead to different results, with or without the $(\sin^2\theta_k)^{-u}$ factor. Since the EEC has linear terms in $k^0$, one must use the equations of motion after the bubble sum, as in \eq{bubble-4d}, because higher-order terms in the series $k^0 = |\vec k| + k^2/(2|\vec k|) + \ldots$ all contribute with equal weight to the renormalon. This constrasts with offshell terms in ${\cal A}$ like $k^2/(p\cdot\bar p)$, which can safely be dropped (and could only modify the residues of poles $u\ge 3/2$). }
\begin{align}\label{eq:bubble-4d}
	2\!\int \!\! \frac{d^4k}{2\pi}\,
    \Theta(k^0)\,\text{Im}&\bigg[\frac{1}{(-k^2-i0)^{1+u}}\bigg] 
   {\cal A}(k^0) 
   = \!\int\! \frac{d k^+d k^-}{2\pi} \Theta(k^0) {\cal A}(k^0)
  \,\text{Im}\bigg[ 
   \frac{d^2k_\perp}{(\vec k_\perp^2 -k^+k^--i0)^{1+u}}\bigg]
  \nonumber\\
	& = \frac{\sin(\pi u) }{\pi u}\,  2\pi\!
    \int\! \frac{d k^+d k^-}{4}\, \Theta(k^+)\Theta(k^-)\, 
   (k^+k^-)^{-u} {\cal A}\Big(\frac{k^++k^-}{2}\Big) \, \nonumber\\
	&=  \frac{\sin(\pi u)  }{\pi u}  
     \int\!\! \frac{d^3k}{2\big|\vec k\big|^{1+2u}} (\sin^2\theta_k)^{-u}
   {\cal A}(|\vec k|) 
  \,.
\end{align}
In the last line, we make the change of variable $k^\pm = |\vec k| \pm k^z$ to transform the result back into the form of a standard phase-space integral, and use $k^+k^- = |\vec k_\perp|^2 = |\vec k|^2 \sin^2\theta_k$. 
Here, $\theta_k$ is a polar angle defined relative to any fixed axis $\hat{l}$, a freedom we use later on.
We now have all the ingredients we need for the bubble-approximated cross-section \eq{eec-bubs-1}.

\subsubsection{Borel-space result}

We now insert the LO matrix element \eq{no-bubble-squared}, 
the Borel-transformed bubble propagator \eq{new-propagator-borel}, 
and the phase-space integral relation \eq{bubble-4d} 
into the bubble-modified EEC cross-section \eq{eec-bubs-1},
and find
\begin{align}\label{eq:eec-bubs2}
	\frac{1}{\sigma_0}\frac{d\hat\Sigma_\bub}{dz}
	&=\frac{8}{\pi^2}
	\frac{C_F ( \mu^2 e^c)^u}{Q^4 \beta_0}\, 
	\frac{\sin(u\pi)}{u\pi} \,\, \int d^3p\,d^3\pbar\,d^3k
	\,\, \delta^4(q-{\textstyle\sum} p_i) 
	\,\, (\sin^{2} \theta_{k} )^{-u}
	 \\*
	& \qquad \qquad \qquad \qquad \qquad \qquad \times
	 \sum_{i,j} \frac{E_i E_j}{E\Ebar E_g^{1+2u}}
	\delta(z-z_{ij})\frac{E^2 + \Ebar^2}{(Q-2E)(Q-2\Ebar)}\nonumber
   \,.
\end{align}
We can break this integral up into two pieces. The first term comes from the choice $ij = q\qbar$. 
The second term comes from $ij = qg$ and $\qbar g$, which are equal by symmetry of \eq{eec-bubs2}. 
We illustrate how to evaluate these integrals using the $qg$ case. 
First, we integrate over $d^3\pbar$ using $\delta^3(\vec{q}-\sum \vec{p}\,)$. 
Without loss of generality, we can take the quark along the $\hat{l}$-axis, so  $\theta_k = \theta_{qg}$,  and we can write $d^3p d^3k= 4\pi E^2 E_g^2 dE dE_g d\cos\theta_k d\Delta\phi$, where $\Delta\phi$ is the difference in azimuthal angle between the quark and the gluon. 
We carry out the angular integrals using $\delta(z-z_{qg})$. Next, we eliminate the remaining energy-conserving delta function using $d\Ebar$. This leaves us with a single remaining integral in $E = Qx/2$.

Combining both the $q\qbar$ and $qg/\qbar g$ terms, we have
\begin{align}\label{eq:bub-x-integral}
 \frac{1}{\sigma_0} \frac{d\hat\Sigma_\bub}{dz}
 &= \frac{C_F ( \mu^2 e^c)^u }{\beta_0 Q^{2u}}\frac{\sin(u\pi)}{u\pi}\frac{1}{[z(1-z)]^{1+u}}\,
  \int_0^1 dx\, \frac{x(1-x)^{-2u}}{(1-xz)^{4-2u}} \\
	&\quad \times \bigg\{ z (1-x) \Big[1-2x+2x^2-2x^3z+x^4z^2 \Big]\nonumber\\
	&\qquad + 2(1-z)\Big[1-4xz+x^2(1+2z+4z^2)-2x^3z(1+2z) + 2x^4z^2 \Big]
	\bigg\} \,. \nonumber
\end{align}
It is possible to perform these integrals analytically. Writing
\begin{align} \label{eq:bub-cross-sec}
\frac{1}{\sigma_0} \frac{d\Sigma_\bub}{dz}
 &= \frac{C_F ( \mu^2 e^c)^u }{\beta_0 Q^{2u}}\frac{\sin(u\pi)}{u\pi}\frac{1}{[z(1-z)]^{1+u}}
 \Bigl ( \mathcal{I}_{q\bar q} + 2 \, \mathcal{I}_{qg} \Bigr)
\,, \end{align} 
we find
\begin{align}\label{eq:integrated-results}
\mathcal{I}_{q\bar q} \equiv \,\,
& z \, \int_{0}^{1} dx \frac{x(1-x)^{-2u}}{(1-xz)^{4-2u}} (1-x) \Big[1-2x+2x^2-2x^3z+x^4z^2 \Big]
\nonumber \\
= & \frac{4 u^3 (1\!-\!z)^3-18 u^2 (1\!-\!z)^2-u (1\!-\!z) (z^{2} + 7z -32) \!-\!6 (4 \!-\!3z)}
   {6 (u-2) } \, _2F_1(1,4-2 u,5-2 u,z)
 \nonumber \\ & 
  +\frac{-8 u^4 (1\!-\!z)^2 + 8 u^3  (1\!-\!z)(5\!-\!z)+u^2 (2z^{2} + 50 z - 86)
   -u (2z^{2}+15 z - 90) -36}{6 (u-1) (2 u-3) }
 \,,
\nonumber \\
\mathcal{I}_{qg} \equiv  \,\,
& (1-z) \, \int_{0}^{1} dx \,\, 
\frac{x (1-x)^{-2u}} {(1- x z)^{4-2u}} \Big[1-4xz+x^2(1+2z+4z^2)-2x^3z(1+2z) + 2x^4z^2 \Big]
\nonumber \\
=& \frac{4 u^2 (1\!-\!z)^3-2 u (1\!-\!z)^{2} (7\!-\!3z) +(1\!-\!z) \bigl( 4z^{2} - 14 z+13\bigr) }{2 (u-2)} \, _2F_1(1,4-2 u,5-2
   u,z)
 \nonumber \\ &
   +\frac{1}{(u-1) (2 u-3) (2 u-1) } \Bigl[ -8 u^4 (1\!-\!z)^2 + 12 u^3 (1\!-\!z) (3\!-\!2z)  
   - u^2 \left(30 z^2\!-\!88 z\!+\!60\right) 
   \nonumber \\ & \qquad\qquad\qquad \qquad\qquad\qquad
   + u (18 z^{2} - 55z  +42) + 4 z (3-z) -12 \Bigr] 
\,, \end{align}
where ${}_2F_1(a,b,c,x)$ is the standard hypergeometric function.  
These results are only valid in the region $0<z<1$, as we have not included the full set of contributions needed to carry out the calculation at $z=0$ and $z=1$.

\subsection{Renormalons in the EEC}

We can now identify renormalons in the EEC from \eq{bub-cross-sec}. Recall that renormalons are poles on the positive real $u$-axis; given our normalization for the Borel transform, they generally appear at half-integer or integer values of $u$.
We find that the leading renormalon appears at $u=1/2$, associated with a pole in $\mathcal{I}_{qg}$ but not in $\mathcal{I}_{qq}$.
This is consistent with the picture of leading power corrections arising from the limit in which the measured gluon becomes nonperturbative. 
Examining the $u\to 1/2$ limit, we find
\begin{align} \label{eq:z12ren}
   \frac{1}{\sigma_0}\frac{d\hat\Sigma}{dz} \bigg|_{u\to 1/2}
  &= \frac{\Res_{1/2}}{u-1/2} + \ldots 
 \nn\\
  &= - \frac{4 C_F}{\pi\beta_0}\frac{\mu e^{c/2}}{Q}\frac{1}{[z(1-z)]^{3/2}} \frac{1}{u-1/2} + \ldots
\,, \end{align}
where the ellipses denote terms of ${\cal O}\big((u-1/2)^0\big)$. We can compute the size of the ambiguity $\Delta_{1/2}$ in the $\MSbar$ perturbative series $d\hat\Sigma/dz$ associated with this renormalon. Specifically, we evaluate the inverse Borel integral over a contour encircling the pole in complex $u$-space:
\begin{align} \label{eq:u12amb}
  \Delta_{1/2} \bigg( \frac{1}{\sigma_0}\frac{d\hat\Sigma}{d z} \bigg)
  &= \oint_{u=1/2}d u\, \exp\Big[-u \frac{4\pi}{\beta_0\alpha_s(\mu)}\Big]\,  \frac{ \Res_{1/2}}{u-1/2} 
  \nn\\
  &= -\frac{8 i C_F e^{5/6}}{\beta_0}\frac{1}{[z(1-z)]^{3/2}}  \frac{\Lambda_{\rm QCD} }{Q}
\,. \end{align}
This result is valid for $0<z<1$. Note that renormalon ambiguities are always imaginary.

This form for the ambiguity in \eq{u12amb} makes clear its connection to the $\Lambda_{\rm QCD}/Q$ EEC power correction. 
Using $\sin^3\chi =8[z(1-z)]^{3/2}$, we write \eq{KSpower} in terms of $z$:
\begin{align} \label{eq:MSbar-NP}
   \frac{1}{\sigma_{0}} \frac{d \Sigma}{d z} 
  =  \frac{1}{\sigma_{0}} \frac{d \hat \Sigma}{dz} 
  + \frac{1}{2[z(1-z)]^{3/2}}
  \frac{\bar \Omega_1}{Q} 
  \,.
\end{align}
Thus, the $z$-dependence of the leading renormalon in \eq{z12ren} agrees with the leading EEC power corrections derived in \refscite{Korchemsky:1999kt, Belitsky:2001gf}.
Since the left-hand side of \eq{MSbar-NP} is a renormalon-free observable, the renormalon in the perturbative series $d\hat\Sigma/d z$ cancels against the renormalon in the $\MSbar$ definition of the hadronic parameter $\bar\Omega_1$. We can cross-check our result for $\Res_{1/2}$ with the analogous renormalon bubble-chain calculations for thrust in the dijet limit~\cite{Hoang:2007vb,Abbate:2010xh}, which allows us to infer the ambiguity
\begin{align} \label{eq:final-amb}
 \Delta_{1/2} \big( \bar\Omega_1 \big) &=  \Lambda_{\rm QCD} \frac{16 i C_{F} e^{5/6}}{ \beta_{0} }
  \,.
\end{align}
Thus, we confirm the cancellation of ambiguities between the two terms in \eq{final-amb}, $\Delta_{1/2}(d\Sigma/d z) = 0$.  This cross-check also provides a test of the universality of power corrections between thrust and the EEC, though we caution that the renormalon based calculation does not account for potential ${\cal O}(1)$ hadron-mass corrections which can potentially spoil the universality.

The EEC observable obeys certain sum rules~\cite{Dixon:2019uzg,Kolouglu:2021lig,Korchemsky:2019nzm}.  A particularly simple one, which holds nonperturbatively,\footnote{Note that results for higher $z$ and $(1-z)$ moments of $d\Sigma/dz$ rely on manipulations with massless four-vectors and hence can have nonperturbative corrections in QCD.} states that 
\begin{align} \label{eq:sumrule0}
  \int_0^1\! d z  \frac{d\Sigma}{d z} = \sigma  \,.
\end{align}
Equation~(\ref{eq:sumrule0}) follows from energy conservation, $\sum_i E_i = Q$, and reflects that $d\Sigma/d z$ integrates to the total hadronic cross-section $\sigma=\sigma(e^+e^-\to X)$, just like a regular differential distribution. 
This cross-section only has nonperturbative corrections at ${\cal O}(\Lambda_{\rm QCD}^4/Q^4)$ for massless quarks, implying that to properly treat the $z=0$ and $z=1$ endpoints, we must modify the function $[z(1-z)]^{-3/2}$ such that it integrates to zero. 
The function remains the same for $0<z<1$, suggesting we need a plus distribution. Because we are dealing with two singularities, at $z=0$ and $z=1$, we cannot determine a functional form uniquely by imposing that the integral vanishes. 
The residual ambiguity can be parameterized by a constant $\eta$, where $0<\eta< 1$. 
Constructing standard plus distributions over the intervals $[0,\eta]$ and $[\eta,1]$, and then recombining terms to cancel as much of the $\eta$ dependence as possible, we find that the regulated $z$-dependence for the power correction can be written as
\begin{align} \label{eq:eecplusfn}
  \Big[ [z(1-z)]^{-3/2} \Big]_+ 
  & \equiv 
   \lim_{\epsilon\to 0} \Big\{ \frac{d}{d z} \theta(z-\epsilon)\theta(1-z-\epsilon) G(z) \Big\}
  + \big[\delta(1-z) - \delta(z)\big]  G(\eta)
 \,,
\end{align} 
where $G(z)=-G(1-z)$, and
\begin{align}
  G(z) = \int^z\!\! d z'\: [z'(1-z')]^{-3/2} = \frac{2z-2(1-z)}{z^{1/2} (1-z)^{1/2}} \,.
\end{align}
It is easy to confirm that $\int_0^1 d z\: \big[ [z(1-z)]^{-3/2} \big]_+ = 0$ for any value of $\eta$. 
A natural choice is to require that the plus distribution be symmetric under $z\to 1-z$, implying $\eta=1/2$ because $G(1/2)=0$. In this case, the $G(\eta)$ term in \eq{eecplusfn} would vanish.  With this modification the formula for the EEC in $\MSbar$ with its leading power correction now reads
\begin{align} \label{eq:MSbar-NP-final}
   \frac{1}{\sigma_{0}} \frac{d \Sigma}{d z} 
  =  \frac{1}{\sigma_{0}} \frac{d \hat \Sigma}{dz} 
  + \frac{1}{2 \big[[z(1-z)]^{3/2}\big]_+}
  \frac{\bar \Omega_1}{Q} 
  \,.
\end{align}

\begin{figure}
	\begin{center}
		\captionsetup[subfigure]{oneside,margin={1.5cm,0cm}}
		\begin{subfigure}{0.51\textwidth} \includegraphics[width =\textwidth]{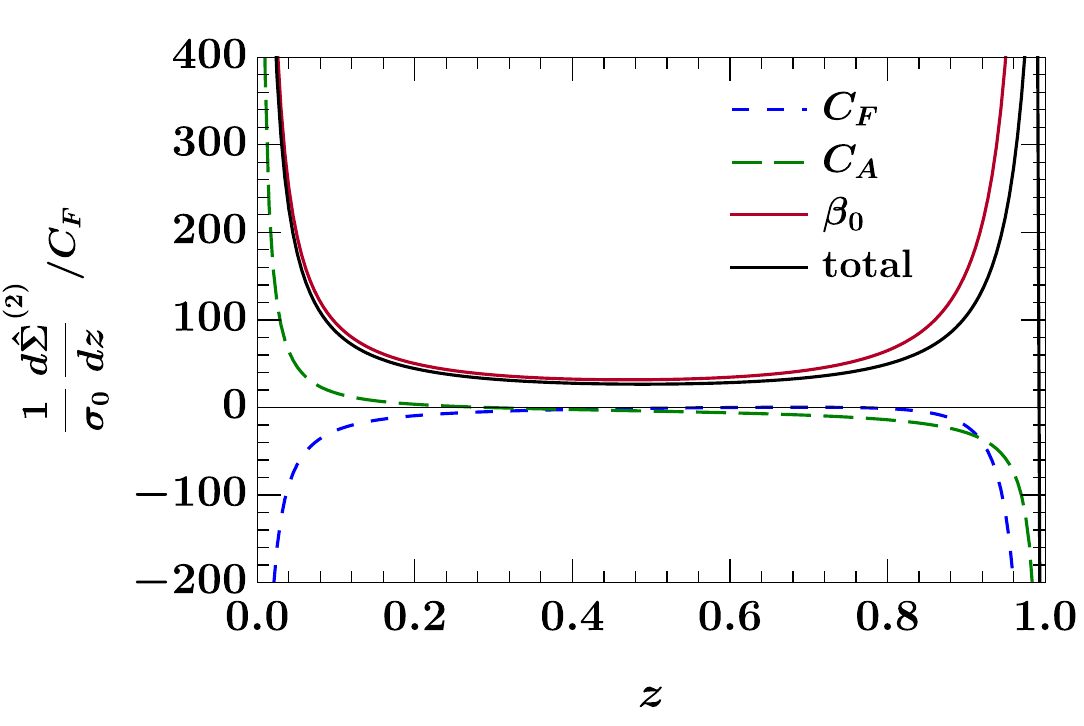}
		\caption{}\label{fig:contributions}\end{subfigure}
		\captionsetup[subfigure]{oneside,margin={1.1cm,0cm}}
		\begin{subfigure}{0.48\textwidth}\includegraphics[width =\textwidth]{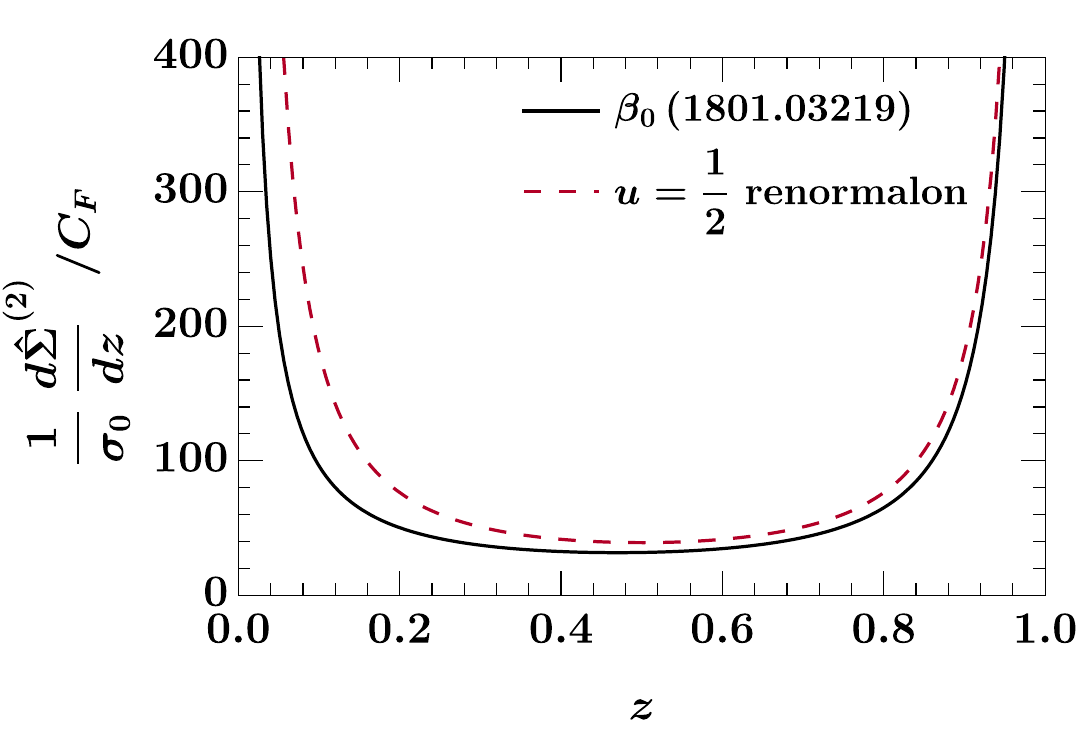}
		\caption{}\label{fig:comparison}\end{subfigure}
	\end{center}
\vspace{-0.3cm}
	\caption{The $\cO(\alpha_s^2)$ (NLO) term in the perturbative series for the EEC, as indicated by the subscript $(2)$ on $d\hat\Sigma^{(2)}$. The normalization used to define this term is $\frac{d\hat\Sigma}{dz} = 2 \big( \frac{\alpha_s}{2\pi} \big) \frac{d\hat\Sigma^{(1)}}{dz} + 2 \big( \frac{\alpha_s}{2\pi}\big)^{2} \frac{d\hat\Sigma^{(2)}}{dz} + ...$. We take $\mu = Q e^{-5/6}$. Figure (a) shows the results calculated in \refcite{Dixon:2018qgp}, organized by color channels $C_F^2$, $C_FC_A$, and $C_F\beta_0$,  as discussed in \sec{bubbles}. The $C_F\beta_0$ term (red) dominates over the other channels. Figure (b) compares the perturbative expansion of the $u=1/2$ renormalon  in \eq{nlo-renormalon} (dashed red) and the $C_F \beta_{0}$ term of the EEC in the $\MSbar$ scheme given in \refcite{Dixon:2018qgp} (solid black) at $\cO(\alpha_s^2)$.}
\end{figure}

While both  $\mathcal{I}_{q \bar q}$ and $\mathcal{I}_{gq}$ exhibit simple poles at $u=1$, the prefactor $\sin(u\pi)$ vanishes here. Thus, the full $d\hat\Sigma_\bub/dz$ has no $u=1$ pole,
\begin{align}
	& \frac{1}{\sigma_0} \frac{d\hat \Sigma}{dz} \bigg|_{u=1} = \,\, \frac{0}{u-1} + \ldots 
\,, \end{align}
where the ellipses are ${\cal O}\big((u-1)^0\big)$. 
This means our bubble-sum analysis has no sensitivity to renormalons corresponding to $\Lambda_{\rm QCD}^2/Q^2$ EEC power corrections.  
We do not, however, rule out a non-Abelian contribution to the $u=1$ residue.
The presence of $u=1$ poles for event shape calculations which treat power corrections from the hard and collinear regions at higher orders in perturbation theory have been discussed in the literature, see for example~\cite{Gardi:2001ny,Gracia:2021nut}.

Our Borel-space result in \eq{bub-cross-sec} indeed has no poles at any integer values of $u=k$, corresponding to $ \Lambda_{\rm QCD}^{2k}/Q^{2k}$ power corrections. We do observe poles at higher half-integer values, $u=k+1/2$, associated with renormalons in $\Lambda_{\rm QCD}^{2k+1}/Q^{2k+1}$ power corrections. Both the $\mathcal{I}_{q \bar q}$ and $\mathcal{I}_{qg}$ terms contribute; for example, near $u=3/2$, we find
\begin{align}
 & \frac{1}{\sigma_0} \frac{d\hat \Sigma}{dz} \bigg|_{u=3/2} 
 = \frac{C_F}{\pi\beta_0}\frac{\mu^3 e^{3c/2}}{Q^3}\frac{(2 - 5 z + 4 z^2)}{[z(1-z)]^{3/2}} \frac{1}{u-3/2} + \ldots
\,, \end{align} 
where the ellipses are ${\cal O}\big((u-3/2)^0\big)$. 
Note that our calculational procedure renders the residues of $u\ge 3/2$ poles somewhat ambiguous because we used $k^2=0$ to simplify the LO matrix element in \eq{no-bubble-squared}. If we had instead expanded around $k^2=0$, the higher-order terms ${\cal O}(k^{2j} / Q^{2j})$ would modify residues at $u\ge 3/2$ (but not at $u=1/2$ or $u=1$).

\subsubsection{Renormalon series in $\MSbar$}

Next, we investigate the manifesation of renormalons in $\MSbar$ perturbative calculations. 
Recall from \fig{contributions} that the dominant contribution to the EEC comes from the $C_F\beta_0$ color structure at ${\cO}(\alpha_s^2)$ \cite{Dixon:2018qgp}.  We determine the extent to which the $u=1/2$ pole approximates the full $C_F\beta_0$ contribution by expanding about $u=0$. 
To retain the leading logarithmic $\mu$-independence when expanding in $u$, we replace $\Res_{1/2} \to  -\frac{1}{\beta_0} {\cal Z} \exp[ 2u \ln(\mu e^{c/2}/Q)]$ (which simplifies back to $\Res_{1/2} = -\frac{1}{\beta_0} {\cal Z} \mu e^{c/2}/Q$ at $u\!\!=\!\!1/2$), and we also expand this exponential. Here ${\cal Z} = 4C_F [z(1-z)]^{-3/2}/\pi$. This gives
\begin{align}\label{eq:expand1}
	\left .\frac{1}{\sigma_{0}}\frac{d\hat\Sigma}{dz} (u) \right\vert_\mathrm{ren.}
	&=   \frac{\Res_{1/2}}{u-1/2}
 \approx  \frac{{\cal Z}}{\beta_0}  \Big[2+4u \big(1+L_Q\big) + 8u^2 \big(1+L_Q+\frac12 L_Q^2\big)  + ... \Big]
\,,\end{align}
where $L_Q = \ln(\mu e^{c/2}/Q) $.
Using the inverse Borel transform in \eq{inverse-borel}, 
\begin{align} \label{eq:nlo-renormalon}
\left .\frac{1}{\sigma_{0}}\frac{d\hat\Sigma}{dz}  (\alpha_s) \right\vert_\mathrm{ren.} \!\!\!\!
&= \frac{{\cal Z}}{\beta_0}\: \sum_{n=1}^{\infty} 2^{n}\, \Gamma(n) \bigg(\frac{\alpha_s(\mu)\beta_0}{4\pi}\bigg)^{\!\!n}
  \ \ \sum_{k=0}^{n-1} \frac{L_Q^k}{k!}
 \\
&
= \frac{{\cal Z}}{\beta_0}\:  \Bigg[ 2 \frac{\alpha_s(\mu)\beta_0}{4\pi} + 4 \bigg(\frac{\alpha_s(
\mu)\beta_0}{4\pi}\bigg)^2 (1+L_Q) + \dots \Bigg] 
\nn .
\end{align}

In \fig{comparison}, we compare the perturbative $C_F\beta_0 \alpha_s^2$ result from \refcite{Dixon:2018qgp} to the $\cO(\alpha_s^2)$ renormalon term in \eq{nlo-renormalon}, taking $\mu=Q e^{-c/2}$ so that $L_Q=0$. We see that the renormalon prediction  mirrors the full $C_F\beta_0$ perturbative result in a substantial portion of $z$-space. This evidentiates the importance of taming terms in \eq{nlo-renormalon} at higher orders, which grow as $2^n (n-1)!$\,.

\section{Mitigating renormalon effects with an R scheme}\label{sec:msr}

In this section, we reorganize the perturbative and nonperturbative corrections to EEC, to remove leading renormalon effects from both of these terms. This leads to improved perturbative convergence and a better agreement with experimental data.

\subsection{EEC in the R scheme}

It is possible to systematically mitigate renormalon effects by switching  from the $\MSbar$ scheme to a so-called R renormalization scheme~\cite{Hoang:2009yr, Hoang:2008yj}. This scheme change removes the dominant Borel pole from the perturbative series, and also modifies matrix elements like $\bar\Omega_1$. The full perturbative series for the EEC in the $\MSbar$ scheme is 
\begin{align}\label{eq:msr1}
	\frac{1}{\sigma_0}\frac{d\hat\Sigma}{dz} 
  =  \sum_{n=1}^\infty c_n(z,\mu/Q) \left[ \frac{\alpha_s(\mu)}{4\pi}\right]^n
 \,,
\end{align}
where $c_n$ is a function of $z$ and $\mu/Q$, \eq{lo-value} gives $c_1$, and \refcite{Dixon:2018qgp} gives $c_2$ analytically. 
To remove the $u=1/2$ renormalon, we define an R scheme using a new subtraction scale $R$. In the R scheme, the leading power correction is~\cite{Hoang:2008fs,Abbate:2010xh}
\begin{align} \label{eq:O1R}
  \Omega_1(R) = \bar\Omega_1 - \bar\delta(R)
  \,.
\end{align}
Here $\bar\delta(R)$ is a perturbative series that removes the leading renormalon,
\begin{align}
  \bar\delta(R) 
 &= R \sum_{n=1}^\infty d_{n0} \left[ \frac{\alpha_s(R)}{4\pi}\right]^n
 = R \sum_{n=1}^\infty d_{n}(\mu/R) 
    \left[ \frac{\alpha_s(\mu)}{4\pi}\right]^n 
 \,,
\end{align}
where $d_{n0}$ are numbers and $d_n(\mu/R)$ are simple functions  $d_n(\mu/R) = \sum_{j=0}^{n-1} d_{nj} \ln^j(\mu/R)$. For $j\ge 1$, $d_{nj}=(2/j) \sum_{k=j}^{n-1} k\, d_{k(j-1)} \beta_{n-k-1}$, where the QCD $\beta$-function coefficients $\beta_n$ are defined by  $\mu \frac{d}{d\mu} \alpha_s(\mu)=-2\alpha_s \sum_{n=0}^\infty \beta_n \alpha_s^{n+1}/(4\pi)^{n+1}$. To ensure that the scheme change in \eq{O1R} does not modify the parametric size of the power correction $\bar\Omega_1$, we choose $R$ such that $\bar\delta(R)\sim \Lambda_{\rm QCD}$. A typical choice is $R\simeq 2\,{\rm GeV}$. In the bubble-sum approximation, the coefficients at large $n$ grow due to the $u=1/2$ renormalon as $d_{n+1}(\mu/R) \simeq (\mu/R) n! (2\beta_0)^n Z$, for constant $Z$, in agreement with \eq{nlo-renormalon}. 

There are multiple possible choices for the $d_{n0}$ coefficients that remove the $u=1/2$ renormalon; \refscite{Hoang:2007vb,Hoang:2008fs,Abbate:2010xh,Bachu:2020nqn} introduce suitable schemes using the hemisphere or thrust soft functions. 
We follow the definition of \refcite{Bachu:2020nqn}, whose coefficients $d_{n0}$ stem from perturbatively expanding the logarithm of the position-space thrust soft function, $\ln \tilde S_{\tau_2}(y,\mu)$, and evaluating it at $i y = 1/\mu = 1/R$. 
Universality implies that this perturbative series has the desired renormalon, and this scheme choice does not rely on the bubble-sum approximation since it contains terms with all allowed color structures. 
Numerically, the first two coefficients in the scheme of \refcite{Bachu:2020nqn} are
\begin{align} \label{eq:dcoeff}
  d_1(\mu/R) &= d_{10} = -8.357 \,,
\\
  d_2(\mu/R) &= d_{20}+ 2\beta_0\, d_{10} \ln\Big(\frac{\mu}{R}\Big) 
       = -72.443 - 16.713 \beta_0  \ln\Big(\frac{\mu}{R}\Big) 
 \nn\,.
\end{align}
Calculating $d_{30}$ requires the non-logarithmic term in the ${\cal O}(\alpha_s^3)$ thrust soft function, for which only estimates exist~\cite{Bruser:2018rad}.  Note that the $d_i(\mu/R)$ coefficients are independent of hadron masses and are therefore suitable for removing the renormalon from $\bar\Omega_1$ for all observables related by massless universality. 

We must also change the EEC perturbative series to the R scheme:
\begin{align}\label{eq:msr2}
	\frac{1}{\sigma_0}\frac{d\hat\Sigma^{\rm R}(R)}{dz} 
&= \sum_{n=1}^\infty \bigg\{ c_n(z,\mu/Q) + \frac{R}{2Q} \frac{d_n(\mu/R)}{\big[[z(1-z)]^{3/2}\big]_+} \bigg\} \left[ \frac{\alpha_s(\mu)}{4\pi}\right]^n
 \,.
\end{align}
Here, the $u=1/2$ renormalon cancels order-by-order between the $c_n$ and $d_n$ series. 
This perturbative R scheme result depends both on the use of the standard $\MSbar$ scheme for the coupling $\alpha_s(\mu)$, and on the R scheme for the power correction.
When carrying out the R scheme change on the full EEC observable, we have:
\begin{align} \label{eq:EEC-MSR}
   \frac{1}{\sigma_{0}} \frac{d \Sigma}{d z} 
  =  \frac{1}{\sigma_{0}} \frac{d \hat \Sigma^{\rm R}(R)}{dz} 
  + \frac{1}{2\big[ [z(1-z)]^{3/2}\big]_+}
  \frac{\Omega_1(R)}{Q} 
  \,.
\end{align}
This represents an improvement over the $\MSbar$ EEC in \eq{MSbar-NP}: here, neither the perturbative series in the first term nor the power correction in the second term has a $u\!\!=\!\!1/2$ renormalon.\footnote{One can conceptualize changing to an R scheme as locating a resurgence-pair of ambiguities in the perturbative and nonperturbative sectors of a trans-series, and shuffling the ambiguities so they cancel out. }
Notably, this renormalon removal relies on the predicted $z$-dependence of the power correction, but {\em does not} rely on the bubble-sum approximation of the residue. 
Indeed, it captures and subtracts even non-Abelian contributions to the renormalon.

\subsection{R scheme with resummation}

Equation~(\ref{eq:EEC-MSR}) still has one remaining issue, associated with large logarithms induced by the scale $R$. To avoid large logarithms in $c_n(z,\mu/Q)$ and $d_n(\mu/R)$, we see from \eq{msr2} that we would need $R\simeq\mu\simeq Q$. Retaining the scaling $\Omega_1(R) \sim \Lambda_{\rm QCD}$, however, requires a relatively small value of $R$, like $R\simeq 2\,{\rm GeV}$; and this scale would cause large $\ln(\mu/R)\simeq \ln(Q/R)$ in $d_n(\mu/R)$. We can resolve these conflicting needs by using the RGE for $R$ in \refcite{Hoang:2008fs}, which has been implemented for $\Omega_1(R)$ in \refscite{Abbate:2010xh,Bachu:2020nqn},
\begin{align}
  \Omega_1(R_1) = \Omega_1(R_0) + K(R_1,R_0)
   = \Omega_1(R_0) -  \sum_{n=0}^\infty \gamma_n^{\Omega_1,R} \int_{R_0}^{R_1} dR \Big[\frac{\alpha_s(R)}{4\pi}\Big]^{n+1} 
  \,.
\end{align} 
Here $K(R_1,R_0)$ is a dimension-1 evolution kernel that sums large logarithms between $R_0$ and $R_1$.
For leading logarithmic (LL) resummation, we need the anomalous dimension $\gamma_0^{\Omega_1,R}=d_{10}=-8.357$,
while at next-to-leading logarithmic (NLL), we also need 
$\gamma_1^{\Omega_1,R}=d_{20}-2\beta_0 d_{10}=55.693$.
Here, we again quote numerical values from \refcite{Bachu:2020nqn}, with $n_f=5$ active light flavors, suitable for the $Q=m_Z$ that we use in our numerical analysis.
Calculating $\gamma_2^{\Omega_1,R}$ requires $d_{30}$ as input, which is not yet available.

This solution to the $R$-RGE enables us to write
\begin{align} \label{eq:EEC-MSR-final}
   \frac{1}{\sigma_{0}} \frac{d \Sigma}{d z} 
  =  \bigg[ \frac{1}{\sigma_{0}} \frac{d \hat \Sigma^{\rm R}(R_1)}{dz} 
  + \frac{K(R_1,R_0)}{2 Q \big[ [z(1-z)]^{3/2}\big]_+} 
  \bigg]
  + \frac{\Omega_1(R_0)}{2 Q\big[ [z(1-z)]^{3/2}\big]_+} 
  \,,
\end{align}
where the first term in square brackets corresponds to the resummed perturbative prediction for the EEC, while the last term contains the R scheme power correction. Here, $R_0\simeq 2\,{\rm GeV}$ and $R_1\simeq Q$. 

\begin{figure}[t!]
	\begin{center} 
   \phantom{x}\hspace{-0.3cm} 
	\includegraphics[width = 0.47\textwidth]{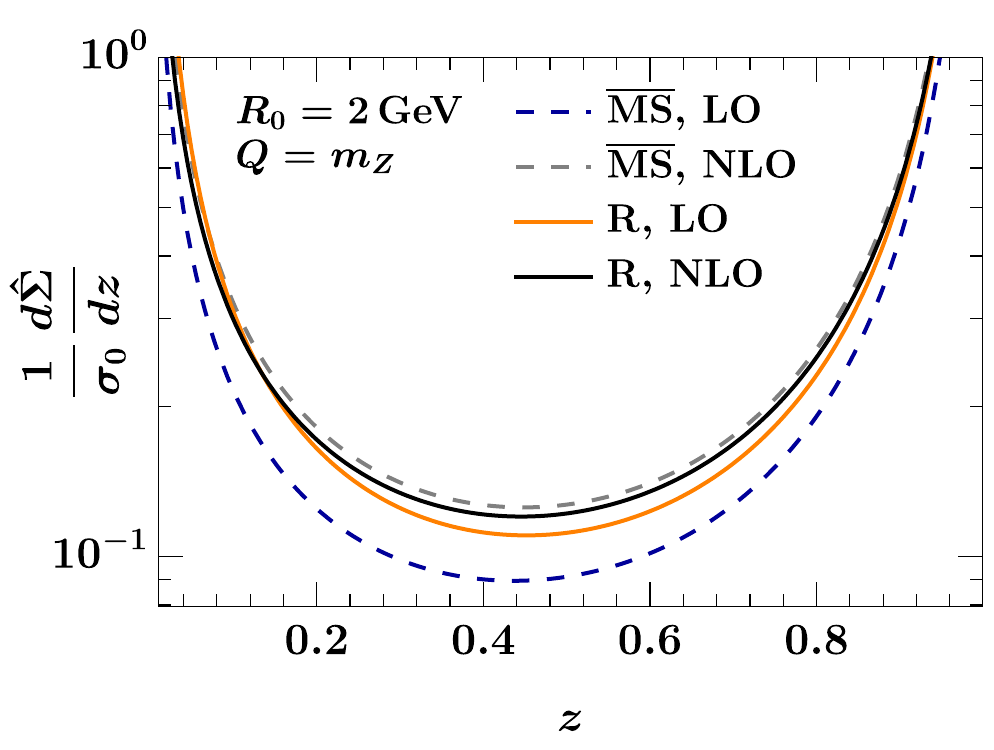} 
	\includegraphics[width = 0.52\textwidth]{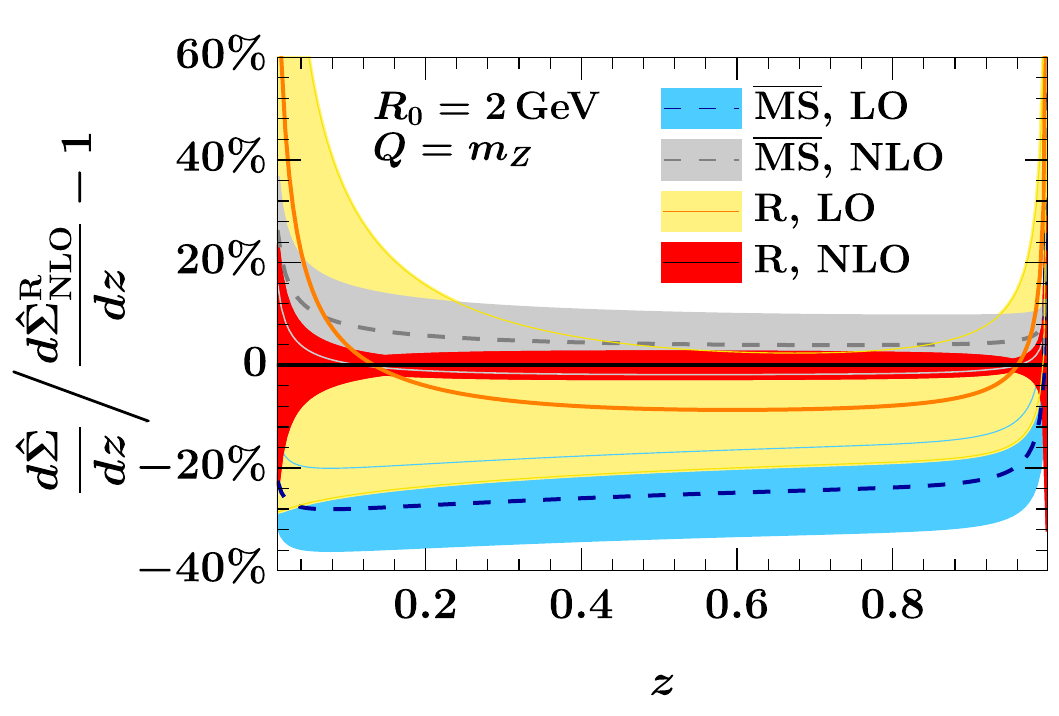}
	\end{center}
\vspace{-0.6cm}
	\caption{Comparison of perturbative results for the  EEC cross-section in the $\MSbar$ and  R schemes in \eqs{final-amb}{EEC-MSR}, respectively. Here, LO is ${\cal O}(\alpha_s)$ and NLO is ${\cal O}(\alpha_s^2)$. 
    The R scheme results have overlapping uncertainty bands and better convergence than $\MSbar$.  
 }\label{fig:MSR-resum}
\end{figure} 

\subsection{Perturbative convergence}

Next, we compute the R scheme perturbative EEC cross-section $d\hat\Sigma^{\rm R}/dz$ at LO and NLO from \eq{msr2}. For simplicity, we take $\mu=R_1$ in $d\hat\Sigma^{\rm R}/dz$.   At LO, we carry out the resummation in $K(R_1,R_0)$ at NLL using $\gamma_{0,1}^{\Omega_1,R}$. At NLO, one should use $K$ at NNLL order; however, $\gamma_2^{\Omega_1,R}$ is not yet known, so we use $K$ at NLL. Based on previous R-RGE studies~\cite{Hoang:2009yr, Hoang:2008yj,Hoang:2017suc}, we anticipate that the NNLL kernel would lead to only slightly smaller perturbative uncertainties at NLO. 

In \fig{MSR-resum}, we set $Q=m_Z$ and $\alpha_s(m_Z)=0.118$, and use $R_0=2\,{\rm GeV}$ for the R scheme.\footnote{A useful feature of R schemes is that the cutoff $R_0$ dependence cancels between the perturbative result and the $\Omega_1$ power correction in \eq{EEC-MSR-final}.  If we leave out $\Omega_1$, we can estimate the size of this power correction by varying $R_0$ in the perturbative term. We do not make use of this method here.}
We assess the uncertainty from higher-order perturbative corrections in the conventional manner, by varying the renormalization scale dependences $\mu$ in $\MSbar$ and $R_1=\mu$ in R scheme, both on the range $Q/2$ to $2Q$, using the LL and NLL running of $\alpha_s(\mu)$ at LO and NLO, respectively. 
In the left panel of \fig{MSR-resum}, we show central values; in the right panel, we show deviation from the NLO R scheme curve and also include perturbative uncertainties. 
The $\MSbar$ EEC value increases significantly from LO to NLO, and the NLO curve is not within the estimated LO uncertainty. In contrast, the R scheme exhibits improved convergence, with NLO results contained within the LO uncertainty band.  Furthermore,  the estimated perturbative uncertainty at NLO is about a factor of two smaller for the R scheme than for $\MSbar$.
Note that (numerical) results exist for the EEC at NNLO=${\cal O}(\alpha_s^3)$ in the $\MSbar$ scheme~\cite{DelDuca:2016csb, Tulipant:2017ybb}, and these results lie just above the upper edge of the NLO $\MSbar$ uncertainty band for almost all values of $z$.
We do not analyze these results here as corresponding NNLO R scheme results are not yet available. 

Note that so far, we have only considered the EEC perturbative series, i.e. the first $\MSbar$ term in \eq{MSbar-NP-final} and the bracketed R scheme terms in \eq{EEC-MSR-final}.
We do not expect these series to asymptote to the same value, as the nonperturbative parameters $\bar{\Omega}_{1}$ and $\Omega_{1}(R_{0})$ differ.  

\subsection{Nonperturbative corrections and hadron masses}\label{sec:np-mass}

To make a more complete prediction for the EEC that includes nonperturbative information, we capitalize on the universality between the thrust and EEC power corrections. 
\Refcite{Abbate:2010xh} determined the  thrust power corrections $\bar\Omega_1$ and $\Omega_1(R_0)$ by fits to data. 
These fits also yielded $\alpha_s(m_Z)=0.114$, which for consistency we use throughout this section, along with $R_0=2\,{\rm GeV}$.
At N$^3$LL$^\prime+{\cal O}(\alpha_s^3)$ order, they obtained $\bar\Omega_1 = 0.252\pm 0.069\, {\rm GeV}$ and $\Omega_1^{\mathrm{ref.}[51]}(R_0)=0.323\pm 0.045$, which translates to $\Omega_1(R_0)=0.739 \pm 0.045{\rm GeV}$ in the R scheme of \refcite{Bachu:2020nqn} used here.\footnote{Compared to \eq{dcoeff}, the scheme of \refcite{Abbate:2010xh} has $d_{10}=0$ and non-zero results only at higher orders, which is the main cause of this numerical difference.}

If we treat all hadrons as massless, then universality implies that $\bar\Omega_1$ and $\Omega_1(R)$ are the same for both thrust and the EEC \cite{Korchemsky:1999kt,Belitsky:2001gf,Lee:2006nr}. Since $\Omega_1$ is the vacuum matrix element of a measurement operator sandwiched by back-to-back lightcone Wilson lines, power corrections for such observables are related by boost symmetry~\cite{Lee:2006nr}. The assumption that hadrons are massless enters such calculations when manipulating kinematic variables to translate between energies, momenta, rapidity, and angles.

\begin{figure}[t!]
	\begin{center} 
	\includegraphics[width = 0.7\textwidth]{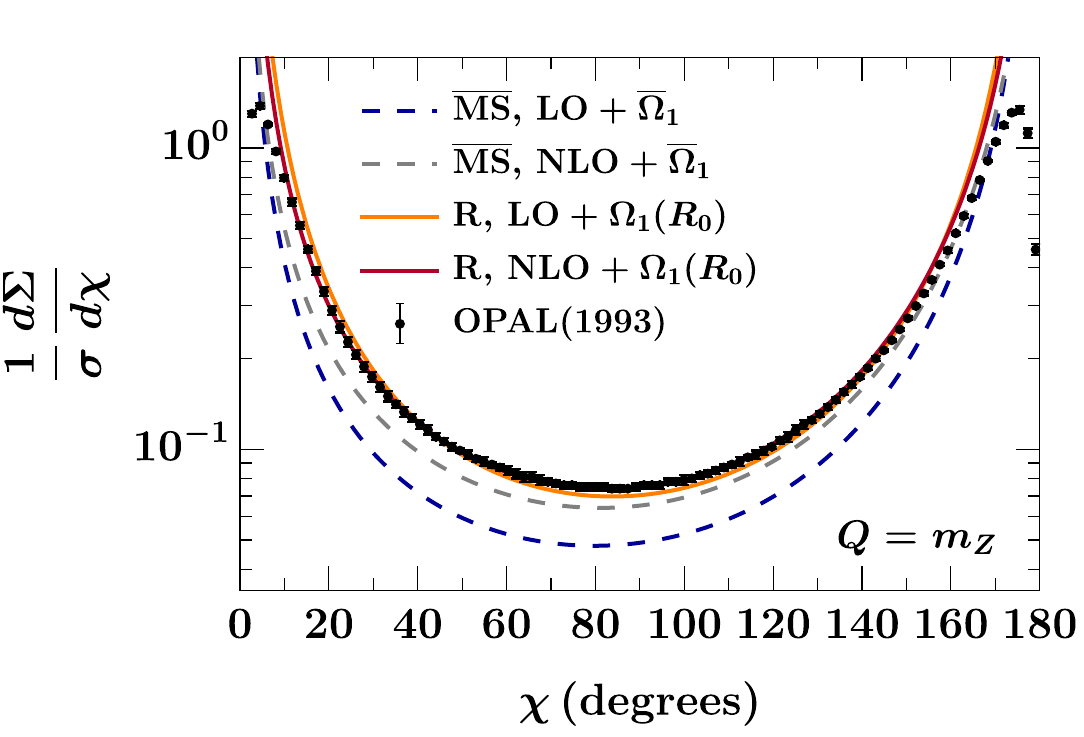}
	\end{center}
\vspace{-0.6cm}
	\caption{Predictions for the full EEC with the inclusion of the leading nonperturbative correction in the $\MSbar$ and R schemes,  compared to OPAL data \cite{OPAL:1993pnw}, as discussed in \sec{np-mass}.
   Central values are displayed at both LO=${\cal O}(\alpha_s)$ and NLO=${\cal O}(\alpha_s^2)$.}
	\label{fig:NP-data-compare}
\end{figure}

Hadron mass corrections are known to have a significant impact on universality relations for $\Omega_1$~\cite{Salam:2001bd}. 
This is influenced by assumptions made about hadron masses when using experimental measurements to determine a given observable.
\Refcite{Mateu:2012nk} 
developed a field-theoretic method to compute hadron mass corrections to universality relations. 
For an observable $e$, we denote the leading-power mass correction by $\Omega_1^e = c_e \Omega_1^{g_e}$, where $\Omega_1^{g_e} = \int_0^1 dr \, g_e(r) \, \Omega_1(r)$ and $r=p_T/\sqrt{p_T^2+m^2}$ for a hadron of transverse momentum $p_T$ and mass $m$. Here $c_e$ and $g_e(r)$ are analytically calculable terms, while $\Omega_1(r)$ is a universal hadronic matrix element. 
Following the notation of \refcite{Mateu:2012nk}, for the EEC we define 
\begin{align}
f_{\rm EEC}(r,y) 
  &= 2 \cosh y\:\delta\Big(\frac{1-\cos\theta}{2} -z\Big) 
   = 2 \cosh y\:\delta\bigg( \frac{1}{2} - \frac{\sinh y}{2\sqrt{r^2+\sinh^2 y}} -z \bigg) \,.
\end{align}
Here, $y$ is rapidity and the prefactor $2$ is the combinatoric factor for two energies. Then,
\begin{align}
 c_{\rm EEC} &= \int_{-\infty}^{+\infty} dy\: f_{\rm EEC}(1,y) =  \frac{1}{2 [z(1-z)]^{3/2} }\,,
\nn\\
 g_{\rm EEC}(r) & = \frac{1}{c_{\rm EEC}} \int_{-\infty}^{+\infty} dy\: f_{\rm EEC}(r,y) = r \,.
\end{align}
Here, $c_{\rm EEC}$ is the coefficient of $\Omega_1$ for the massless universality relation, in agreement with \eqs{MSbar-NP-final}{EEC-MSR-final}. 

The result $g_{\rm EEC}(r)=r$ implies that  $\Omega_1^{g_{\rm EEC}}$ is in the so-called E-scheme universality class of power corrections, which differs from the thrust universality class $\Omega_1^{g_\tau}$ quoted above.
Since $g_{\rm EEC}(r) > g_\tau(r)$, we expect the value of the EEC power correction to be larger than that for thrust. 
A two-term basis expansion provides a fairly accurate parametrization for the impact of hadron masses on $g(r)$~\cite{ Mateu:2012nk}, 
enabling us to write any observable as a linear combination of nonperturbative parameters $\Omega_1^{g_e} = b_0^e \Omega_1^{(0)} + b_1^e \Omega_1^{(1)}$.  
Using the R scheme thrust fit, and $\Omega_1^{(0)}-\Omega_1^{(1)} \simeq 0.7\,{\rm GeV}$ from Monte Carlo fits~\cite{ Mateu:2012nk}, we find $\Omega_1^{g_{\rm EEC}}/\Omega_1^{g_\tau} =1.21$. 
Thus,  hadron mass effects in the EEC induce a 21\% increase in the value of $\Omega_1(R_0)$, giving $\Omega_{1}^{g_\mathrm{EEC}}(R_0) = 0.895 \pm 0.054\, {\rm GeV}$. 
It is harder to obtain an analogous $\MSbar$ estimate, as Monte Carlo simulations with hadronization at the shower cutoff are physically similar to the R scheme, but not to $\MSbar$. 
Thus, we simply assume that $\MSbar$ hadron mass corrections also cause a 21\% increase, yielding
$\bar\Omega_1^{g_\mathrm{EEC}} = 0.305\pm 0.084\, {\rm GeV}$.

\subsection{Comparison to experimental data}\label{sec:data}

\begin{figure}[t!]
	\begin{center} 
	\includegraphics[width = 0.495\textwidth]{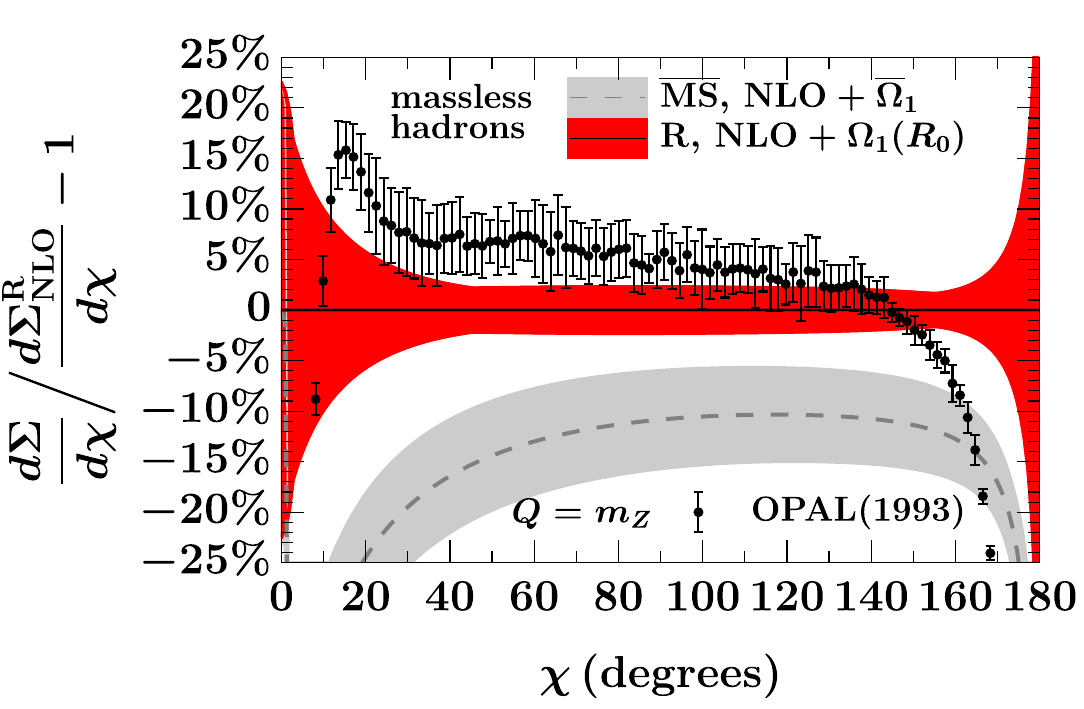}
	\includegraphics[width = 0.495\textwidth]{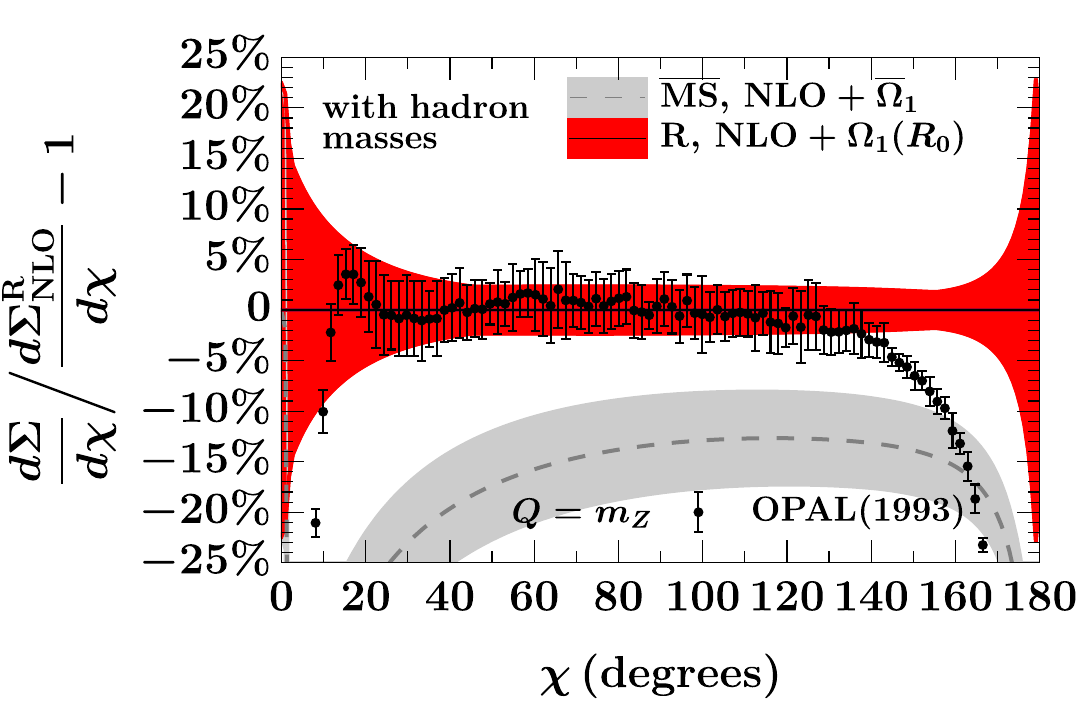}
	\end{center}
	\vspace{-0.6cm}
	\caption{Full nonperturbative predictions for the EEC without (left)  and with (right) corrections from hadron mass effects as discussed in \secs{np-mass}{data}, normalized to NLO R scheme. 
	We show the uncertainty bands for NLO curves with the perturbative and parametric uncertainties added in quadrature.}
	\label{fig:NP-hadron-mass}
\end{figure}

We now combine the perturbative EEC results and leading nonperturbative corrections to predict the total EEC cross-section. 
In \fig{NP-data-compare}, we compare our $\MSbar$ and R scheme results with OPAL data \cite{OPAL:1993pnw}.
Here, $\bar\Omega_1$ and $\Omega_1(R_0)$ include the estimated hadron mass corrections.
We normalize the theory results by dividing by the full hadronic cross-section $\sigma$, to ${\cal O}(\alpha_s)$ and ${\cal O}(\alpha_s^2)$ for LO and NLO~\cite{Chetyrkin:1996ela}.
Our NLO R scheme result agrees well with OPAL data, except in the $z\to 0$ and $z\to 1$ regions, where one should resum large perturbative logarithms to higher orders in $\alpha_s$.  Note that the NLO R scheme result is also closer to the data than the LO R scheme result. For $\MSbar$, the power correction shifts $\MSbar$ predictions closer to the data, but it is apparent that one needs higher-order perturbative corrections to improve agreement.

It is difficult to illustrate hadron mass effects and uncertainties in the theoretical predictions on the scales shown in \fig{NP-data-compare}.
Therefore, we display the same data as percent deviations from the NLO R scheme result in \fig{NP-hadron-mass}.
In the left panel, we use the massless hadron parameter values, corresponding to the thrust fit values of $\bar \Omega_1$ and $\Omega_1(R_0)$.%
\footnote{Though we have only shown predictions for the EEC and not considered fits to the EEC data, we do note that fits at a single value of $Q$ are insufficient to break the degeneracy between different values of $\alpha_{s}(m_{Z})$ and $\Omega_{1}(R_{0})$. For example, had we fixed $\alpha_{s}(m_{Z})=0.118$, we could find a value of $\Omega_{1}(R_{0})$ that gives similarly good agreement with the OPAL data, as in the right panel of \fig{NP-hadron-mass}.} 
In the right panel, we include the 21\% increase in these parameters due to the hadron mass effects discussed in \sec{np-mass}; we find that this strongly improves the agreement between theory and data in both normalization and shape.
We show both the perturbative uncertainties at NLO from varying $\mu$ and $R_{1}$
and the parametric uncertainties from the fit used to obtain values for $\Omega_{1}(R_{0})$ and $\bar\Omega_{1}$,
quoted in \sec{np-mass}.
These two types of uncertainties are added in quadrature in \fig{NP-hadron-mass}.
We note that the $\MSbar$ perturbative uncertainty 
is about twice as large as the parametric uncertainty of $\bar\Omega_{1}$,
and the R scheme perturbative uncertainty 
is of similar size as the parametric uncertainty of $\Omega_{1}(R_{0})$.
We also remark that in all cases the power corrections give improved agreement with data compared to the purely perturbative EEC.  Furthermore, in the R  scheme, the fixed-order result using a single $R=R_0=R_1$ and resummation kernel $K=0$ exhibits worse convergence and worse agreement with data than the NLO R scheme result shown here.

\section{Concluding remarks}\label{sec:conclusion}

Energy correlators show great promise for extracting a breadth of information about QCD from colliders. To fully capitalize on the opportunity presented by this class of observables, one must reach a high level of precision in both perturbative calculations and nonperturbative corrections. 
In this paper, we analyzed renormalons in the $\MSbar$ EEC to investigate the nature of its asymptotic perturbative series, 
and we converted our results to an R scheme that removes the leading renormalon ambiguity,
significantly improving convergence and agreement with experimental data. 
Specifically, we computed the Borel space $\MSbar$ EEC in the bubble-sum approximation, allowing us to calculate the $z$-dependence of the leading $u=1/2$ renormalon pole.
We confirmed the expected $[z(1-z)]^{-3/2}$ dependence of the corresponding $\cO(\Lambda_{\rm QCD}/Q)$ power correction, 
and showed how to regulate the $z\to 0$ and $z\to 1$ endpoints. 
We found no $u=1$ pole under this approximation. 
Next, we constructed an R renormalization scheme that removes the leading $u=1/2$ renormalon from both the perturbative and the nonperturbative contributions to the EEC, with explicit results given to $\cO(\alpha_s^2)$. 
This scheme change significantly improves the convergence of the EEC perturbative series
and brings theoretical calculations of the EEC into reasonable agreement with experimental data already at ${\cal O}(\alpha_s^2)$. 
This paper lays the groundwork for improving theoretical predictions for other EEC observables, such as higher-point EEC correlators (see e.g. \refcite{Chen:2019bpb}) and EEC observables relevant for electron-ion and hadron-hadron colliders.
It would also be interesting to extend the analysis here to make use of the ${\cal O}(\alpha_s^3)$ $\MSbar$ perturbative results for the two-point EEC~\cite{DelDuca:2016csb, Tulipant:2017ybb}.

\section*{Acknowledgments} 

We thank Andre Hoang, Kyle Lee, Johannes Michel, Ian Moult, Michael Ogilvie, and Aditya Pathak for helpful conversations.
This work was supported by the U.S. Department of Energy, Office of Science, Office of Nuclear Physics from DE-SC0011090. 
I.S. was also supported in part by the Simons Foundation through the Investigator grant 327942.
S.T.S. was partially supported by the U.S. National Science Foundation through a Graduate Research Fellowship under Grant No. 1745302, and fellowships from the MIT Department of Physics and MIT School of Science.
Z.S. was also supported by a fellowship from the MIT Department of Physics.

\bibliographystyle{JHEP}
\bibliography{eec-renormalon}

\providecommand{\href}[2]{#2}\begingroup\raggedright\begin{thebibliography}{100}

\bibitem{PhysRevLett.41.1585}
C.~L. Basham, L.~S. Brown, S.~D. Ellis and S.~T. Love, \emph{Energy
  correlations in electron-positron annihilation: Testing quantum
  chromodynamics},
  \href{https://doi.org/10.1103/PhysRevLett.41.1585}{\emph{Phys. Rev. Lett.}
  {\bfseries 41} (1978) 1585}.

\bibitem{PhysRevD.19.2018}
C.~L. Basham, L.~S. Brown, S.~D. Ellis and S.~T. Love, \emph{Energy
  correlations in electron-positron annihilation in quantum chromodynamics:
  Asymptotically free perturbation theory},
  \href{https://doi.org/10.1103/PhysRevD.19.2018}{\emph{Phys. Rev. D}
  {\bfseries 19} (1979) 2018}.

\bibitem{Ali:1984yp}
A.~Ali, E.~Pietarinen and W.~J. Stirling, \emph{{Transverse Energy-energy
  Correlations: A Test of Perturbative {QCD} for the Proton - Anti-proton
  Collider}}, \href{https://doi.org/10.1016/0370-2693(84)90283-1}{\emph{Phys.
  Lett. B} {\bfseries 141} (1984) 447}.

\bibitem{Chen:2019bpb}
H.~Chen, M.-X. Luo, I.~Moult, T.-Z. Yang, X.~Zhang and H.~X. Zhu, \emph{{Three
  point energy correlators in the collinear limit: symmetries, dualities and
  analytic results}},
  \href{https://doi.org/10.1007/JHEP08(2020)028}{\emph{JHEP} {\bfseries 08}
  (2020) 028} [\href{https://arxiv.org/abs/1912.11050}{{\ttfamily
  1912.11050}}].

\bibitem{Li:2021txc}
H.~T. Li, Y.~Makris and I.~Vitev, \emph{{Energy-energy correlators in Deep
  Inelastic Scattering}},
  \href{https://doi.org/10.1103/PhysRevD.103.094005}{\emph{Phys. Rev. D}
  {\bfseries 103} (2021) 094005}
  [\href{https://arxiv.org/abs/2102.05669}{{\ttfamily 2102.05669}}].

\bibitem{Li:2020bub}
H.~T. Li, I.~Vitev and Y.~J. Zhu, \emph{{Transverse-Energy-Energy Correlations
  in Deep Inelastic Scattering}},
  \href{https://doi.org/10.1007/JHEP11(2020)051}{\emph{JHEP} {\bfseries 11}
  (2020) 051} [\href{https://arxiv.org/abs/2006.02437}{{\ttfamily
  2006.02437}}].

\bibitem{Chen:2020vvp}
H.~Chen, I.~Moult, X.~Zhang and H.~X. Zhu, \emph{{Rethinking jets with energy
  correlators: Tracks, resummation, and analytic continuation}},
  \href{https://doi.org/10.1103/PhysRevD.102.054012}{\emph{Phys. Rev. D}
  {\bfseries 102} (2020) 054012}
  [\href{https://arxiv.org/abs/2004.11381}{{\ttfamily 2004.11381}}].

\bibitem{Komiske:2022enw}
P.~T. Komiske, I.~Moult, J.~Thaler and H.~X. Zhu, \emph{{Analyzing N-Point
  Energy Correlators inside Jets with CMS Open Data}},
  \href{https://doi.org/10.1103/PhysRevLett.130.051901}{\emph{Phys. Rev. Lett.}
  {\bfseries 130} (2023) 051901}
  [\href{https://arxiv.org/abs/2201.07800}{{\ttfamily 2201.07800}}].

\bibitem{Lee:2022ige}
K.~Lee, B.~Me\c{c}aj and I.~Moult, \emph{{Conformal Colliders Meet the LHC}},
  \href{https://arxiv.org/abs/2205.03414}{{\ttfamily 2205.03414}}.

\bibitem{Neill:2022lqx}
D.~Neill, G.~Vita, I.~Vitev and H.~X. Zhu, \emph{{Energy-Energy Correlators for
  Precision QCD}},  in \emph{{2022 Snowmass Summer Study}}, 3, 2022,
  \href{https://arxiv.org/abs/2203.07113}{{\ttfamily 2203.07113}}.

\bibitem{Burrows:1995vt}
P.~N. Burrows, H.~Masuda, D.~Muller and Y.~Ohnishi, \emph{{Application of
  'optimized' perturbation theory to determination of alpha-s M(Z)**2 from
  hadronic event shape observables in e+ e- annihilation}},
  \href{https://doi.org/10.1016/0370-2693(96)00570-9}{\emph{Phys. Lett. B}
  {\bfseries 382} (1996) 157}
  [\href{https://arxiv.org/abs/hep-ph/9602210}{{\ttfamily hep-ph/9602210}}].

\bibitem{Kardos:2018kqj}
A.~Kardos, S.~Kluth, G.~Somogyi, Z.~Tulip\'ant and A.~Verbytskyi,
  \emph{{Precise determination of $\alpha _{S}(M_Z)$ from a global fit of
  energy\textendash{}energy correlation to NNLO+NNLL predictions}},
  \href{https://doi.org/10.1140/epjc/s10052-018-5963-1}{\emph{Eur. Phys. J. C}
  {\bfseries 78} (2018) 498}
  [\href{https://arxiv.org/abs/1804.09146}{{\ttfamily 1804.09146}}].

\bibitem{Gao:2019ojf}
A.~Gao, H.~T. Li, I.~Moult and H.~X. Zhu, \emph{{Precision QCD Event Shapes at
  Hadron Colliders: The Transverse Energy-Energy Correlator in the Back-to-Back
  Limit}}, \href{https://doi.org/10.1103/PhysRevLett.123.062001}{\emph{Phys.
  Rev. Lett.} {\bfseries 123} (2019) 062001}
  [\href{https://arxiv.org/abs/1901.04497}{{\ttfamily 1901.04497}}].

\bibitem{OPAL:1990reb}
{\scshape OPAL} collaboration, M.~Z. Akrawy et~al., \emph{{A Measurement of
  energy correlations and a determination of alpha-s (M2 (Z0)) in e+ e-
  annihilations at s**(1/2) = 91-GeV}},
  \href{https://doi.org/10.1016/0370-2693(90)91098-V}{\emph{Phys. Lett. B}
  {\bfseries 252} (1990) 159}.

\bibitem{ALEPH:1990vew}
{\scshape ALEPH} collaboration, D.~Decamp et~al., \emph{{Measurement of alpha-s
  from the structure of particle clusters produced in hadronic Z decays}},
  \href{https://doi.org/10.1016/0370-2693(91)91926-M}{\emph{Phys. Lett. B}
  {\bfseries 257} (1991) 479}.

\bibitem{L3:1991qlf}
{\scshape L3} collaboration, B.~Adeva et~al., \emph{{Determination of alpha-s
  from energy-energy correlations measured on the Z0 resonance.}},
  \href{https://doi.org/10.1016/0370-2693(91)91925-L}{\emph{Phys. Lett. B}
  {\bfseries 257} (1991) 469}.

\bibitem{ATLAS:2015yaa}
{\scshape ATLAS} collaboration, G.~Aad et~al., \emph{{Measurement of transverse
  energy-energy correlations in multi-jet events in $pp$ collisions at
  $\sqrt{s} = 7$ TeV using the ATLAS detector and determination of the strong
  coupling constant $\alpha_{\mathrm{s}}(m_Z)$}},
  \href{https://doi.org/10.1016/j.physletb.2015.09.050}{\emph{Phys. Lett. B}
  {\bfseries 750} (2015) 427}
  [\href{https://arxiv.org/abs/1508.01579}{{\ttfamily 1508.01579}}].

\bibitem{ATLAS:2017qir}
{\scshape ATLAS} collaboration, M.~Aaboud et~al., \emph{{Determination of the
  strong coupling constant $\alpha _\mathrm {s}$ from transverse
  energy\textendash{}energy correlations in multijet events at $\sqrt{s} =
  8~\hbox {TeV}$ using the ATLAS detector}},
  \href{https://doi.org/10.1140/epjc/s10052-017-5442-0}{\emph{Eur. Phys. J. C}
  {\bfseries 77} (2017) 872}
  [\href{https://arxiv.org/abs/1707.02562}{{\ttfamily 1707.02562}}].

\bibitem{ATLAS:2020mee}
{\scshape ATLAS} collaboration, \emph{{Determination of the strong coupling
  constant and test of asymptotic freedom from Transverse Energy-Energy
  Correlations in multijet events at $\sqrt{s} = 13$ TeV with the ATLAS
  detector}}, .

\bibitem{SLD:1994yoe}
{\scshape SLD} collaboration, K.~Abe et~al., \emph{{Measurement of alpha-s from
  energy-energy correlations at the Z0 resonance}},
  \href{https://doi.org/10.1103/PhysRevD.50.5580}{\emph{Phys. Rev. D}
  {\bfseries 50} (1994) 5580}
  [\href{https://arxiv.org/abs/hep-ex/9405006}{{\ttfamily hep-ex/9405006}}].

\bibitem{AbdulKhalek:2021gbh}
R.~Abdul~Khalek et~al., \emph{{Science Requirements and Detector Concepts for
  the Electron-Ion Collider: EIC Yellow Report}},
  \href{https://arxiv.org/abs/2103.05419}{{\ttfamily 2103.05419}}.

\bibitem{Schneider:1983iu}
H.~N. Schneider, G.~Kramer and G.~Schierholz, \emph{{Higher Order {QCD}
  Corrections to the Energy-energy Correlation Function}},
  \href{https://doi.org/10.1007/BF01572173}{\emph{Z. Phys. C} {\bfseries 22}
  (1984) 201}.

\bibitem{Falck:1988gb}
N.~K. Falck and G.~Kramer, \emph{{Theoretical Studies of Energy-energy
  Correlation in $e^+ e^-$ Annihilation}},
  \href{https://doi.org/10.1007/BF01548452}{\emph{Z. Phys. C} {\bfseries 42}
  (1989) 459}.

\bibitem{Glover:1994vz}
E.~W.~N. Glover and M.~R. Sutton, \emph{{The Energy-energy correlation function
  revisited}}, \href{https://doi.org/10.1016/0370-2693(94)01354-F}{\emph{Phys.
  Lett. B} {\bfseries 342} (1995) 375}
  [\href{https://arxiv.org/abs/hep-ph/9410234}{{\ttfamily hep-ph/9410234}}].

\bibitem{Kramer:1996qr}
G.~Kramer and H.~Spiesberger, \emph{{A New calculation of the NLO energy-energy
  correlation function}}, \href{https://doi.org/10.1007/s002880050339}{\emph{Z.
  Phys. C} {\bfseries 73} (1997) 495}
  [\href{https://arxiv.org/abs/hep-ph/9603385}{{\ttfamily hep-ph/9603385}}].

\bibitem{Ali:1982ub}
A.~Ali and F.~Barreiro, \emph{{An O ($\alpha^- s^2$) Calculation of
  Energy-energy Correlation in $e^+ e^-$ Annihilation and Comparison With
  Experimental Data}},
  \href{https://doi.org/10.1016/0370-2693(82)90621-9}{\emph{Phys. Lett. B}
  {\bfseries 118} (1982) 155}.

\bibitem{Ali:1984gzn}
A.~Ali and F.~Barreiro, \emph{{Energy-energy Correlations in $e^+ e^-$
  Annihilation}},
  \href{https://doi.org/10.1016/0550-3213(84)90536-4}{\emph{Nucl. Phys. B}
  {\bfseries 236} (1984) 269}.

\bibitem{Richards:1982te}
D.~G. Richards, W.~J. Stirling and S.~D. Ellis, \emph{{Second Order Corrections
  to the Energy-energy Correlation Function in Quantum Chromodynamics}},
  \href{https://doi.org/10.1016/0370-2693(82)90275-1}{\emph{Phys. Lett. B}
  {\bfseries 119} (1982) 193}.

\bibitem{Richards:1983sr}
D.~G. Richards, W.~J. Stirling and S.~D. Ellis, \emph{{Energy-energy
  Correlations to Second Order in Quantum Chromodynamics}},
  \href{https://doi.org/10.1016/0550-3213(83)90335-8}{\emph{Nucl. Phys. B}
  {\bfseries 229} (1983) 317}.

\bibitem{Catani:1996jh}
S.~Catani and M.~H. Seymour, \emph{{The Dipole formalism for the calculation of
  QCD jet cross-sections at next-to-leading order}},
  \href{https://doi.org/10.1016/0370-2693(96)00425-X}{\emph{Phys. Lett. B}
  {\bfseries 378} (1996) 287}
  [\href{https://arxiv.org/abs/hep-ph/9602277}{{\ttfamily hep-ph/9602277}}].

\bibitem{DelDuca:2016csb}
V.~Del~Duca, C.~Duhr, A.~Kardos, G.~Somogyi and Z.~Tr\'ocs\'anyi,
  \emph{{Three-Jet Production in Electron-Positron Collisions at
  Next-to-Next-to-Leading Order Accuracy}},
  \href{https://doi.org/10.1103/PhysRevLett.117.152004}{\emph{Phys. Rev. Lett.}
  {\bfseries 117} (2016) 152004}
  [\href{https://arxiv.org/abs/1603.08927}{{\ttfamily 1603.08927}}].

\bibitem{Tulipant:2017ybb}
Z.~Tulip\'ant, A.~Kardos and G.~Somogyi, \emph{{Energy\textendash{}energy
  correlation in electron\textendash{}positron annihilation at NNLL + NNLO
  accuracy}}, \href{https://doi.org/10.1140/epjc/s10052-017-5320-9}{\emph{Eur.
  Phys. J. C} {\bfseries 77} (2017) 749}
  [\href{https://arxiv.org/abs/1708.04093}{{\ttfamily 1708.04093}}].

\bibitem{Dixon:2018qgp}
L.~J. Dixon, M.-X. Luo, V.~Shtabovenko, T.-Z. Yang and H.~X. Zhu,
  \emph{{Analytical Computation of Energy-Energy Correlation at Next-to-Leading
  Order in QCD}},
  \href{https://doi.org/10.1103/PhysRevLett.120.102001}{\emph{Phys. Rev. Lett.}
  {\bfseries 120} (2018) 102001}
  [\href{https://arxiv.org/abs/1801.03219}{{\ttfamily 1801.03219}}].

\bibitem{Dixon:2019uzg}
L.~J. Dixon, I.~Moult and H.~X. Zhu, \emph{{Collinear limit of the
  energy-energy correlator}},
  \href{https://doi.org/10.1103/PhysRevD.100.014009}{\emph{Phys. Rev. D}
  {\bfseries 100} (2019) 014009}
  [\href{https://arxiv.org/abs/1905.01310}{{\ttfamily 1905.01310}}].

\bibitem{Korchemsky:2019nzm}
G.~P. Korchemsky, \emph{{Energy correlations in the end-point region}},
  \href{https://doi.org/10.1007/JHEP01(2020)008}{\emph{JHEP} {\bfseries 01}
  (2020) 008} [\href{https://arxiv.org/abs/1905.01444}{{\ttfamily
  1905.01444}}].

\bibitem{Chen:2020uvt}
H.~Chen, T.-Z. Yang, H.~X. Zhu and Y.~J. Zhu, \emph{{Analytic Continuation and
  Reciprocity Relation for Collinear Splitting in QCD}},
  \href{https://doi.org/10.1088/1674-1137/abde2d}{\emph{Chin. Phys. C}
  {\bfseries 45} (2021) 043101}
  [\href{https://arxiv.org/abs/2006.10534}{{\ttfamily 2006.10534}}].

\bibitem{Kodaira:1981nh}
J.~Kodaira and L.~Trentadue, \emph{{Summing Soft Emission in QCD}},
  \href{https://doi.org/10.1016/0370-2693(82)90907-8}{\emph{Phys. Lett. B}
  {\bfseries 112} (1982) 66}.

\bibitem{Kodaira:1982az}
J.~Kodaira and L.~Trentadue, \emph{{Single Logarithm Effects in
  electron-Positron Annihilation}},
  \href{https://doi.org/10.1016/0370-2693(83)91213-3}{\emph{Phys. Lett. B}
  {\bfseries 123} (1983) 335}.

\bibitem{deFlorian:2004mp}
D.~de~Florian and M.~Grazzini, \emph{{The Back-to-back region in e+ e-
  energy-energy correlation}},
  \href{https://doi.org/10.1016/j.nuclphysb.2004.10.051}{\emph{Nucl. Phys. B}
  {\bfseries 704} (2005) 387}
  [\href{https://arxiv.org/abs/hep-ph/0407241}{{\ttfamily hep-ph/0407241}}].

\bibitem{Moult:2018jzp}
I.~Moult and H.~X. Zhu, \emph{{Simplicity from Recoil: The Three-Loop Soft
  Function and Factorization for the Energy-Energy Correlation}},
  \href{https://doi.org/10.1007/JHEP08(2018)160}{\emph{JHEP} {\bfseries 08}
  (2018) 160} [\href{https://arxiv.org/abs/1801.02627}{{\ttfamily
  1801.02627}}].

\bibitem{Ebert:2020sfi}
M.~A. Ebert, B.~Mistlberger and G.~Vita, \emph{{The Energy-Energy Correlation
  in the back-to-back limit at N$^{3}$LO and N$^{3}$LL'}},
  \href{https://doi.org/10.1007/JHEP08(2021)022}{\emph{JHEP} {\bfseries 08}
  (2021) 022} [\href{https://arxiv.org/abs/2012.07859}{{\ttfamily
  2012.07859}}].

\bibitem{Duhr:2022yyp}
C.~Duhr, B.~Mistlberger and G.~Vita, \emph{{Four-Loop Rapidity Anomalous
  Dimension and Event Shapes to Fourth Logarithmic Order}}, {\emph{Phys. Rev.
  Lett.} {\bfseries 129} (2022) 162001}
  [\href{https://arxiv.org/abs/2205.02242}{{\ttfamily 2205.02242}}].

\bibitem{Belitsky:2013xxa}
A.~V. Belitsky, S.~Hohenegger, G.~P. Korchemsky, E.~Sokatchev and A.~Zhiboedov,
  \emph{{From correlation functions to event shapes}},
  \href{https://doi.org/10.1016/j.nuclphysb.2014.04.020}{\emph{Nucl. Phys. B}
  {\bfseries 884} (2014) 305}
  [\href{https://arxiv.org/abs/1309.0769}{{\ttfamily 1309.0769}}].

\bibitem{Belitsky:2013bja}
A.~V. Belitsky, S.~Hohenegger, G.~P. Korchemsky, E.~Sokatchev and A.~Zhiboedov,
  \emph{{Event shapes in $\mathcal{N} = 4$ super-Yang-Mills theory}},
  \href{https://doi.org/10.1016/j.nuclphysb.2014.04.019}{\emph{Nucl. Phys. B}
  {\bfseries 884} (2014) 206}
  [\href{https://arxiv.org/abs/1309.1424}{{\ttfamily 1309.1424}}].

\bibitem{Belitsky:2013ofa}
A.~V. Belitsky, S.~Hohenegger, G.~P. Korchemsky, E.~Sokatchev and A.~Zhiboedov,
  \emph{{Energy-Energy Correlations in N=4 Supersymmetric Yang-Mills Theory}},
  \href{https://doi.org/10.1103/PhysRevLett.112.071601}{\emph{Phys. Rev. Lett.}
  {\bfseries 112} (2014) 071601}
  [\href{https://arxiv.org/abs/1311.6800}{{\ttfamily 1311.6800}}].

\bibitem{Henn:2019gkr}
J.~M. Henn, E.~Sokatchev, K.~Yan and A.~Zhiboedov, \emph{{Energy-energy
  correlation in $N$=4 super Yang-Mills theory at next-to-next-to-leading
  order}}, \href{https://doi.org/10.1103/PhysRevD.100.036010}{\emph{Phys. Rev.
  D} {\bfseries 100} (2019) 036010}
  [\href{https://arxiv.org/abs/1903.05314}{{\ttfamily 1903.05314}}].

\bibitem{Moult:2019vou}
I.~Moult, G.~Vita and K.~Yan, \emph{{Subleading power resummation of rapidity
  logarithms: the energy-energy correlator in $ \mathcal{N} $ = 4 SYM}},
  \href{https://doi.org/10.1007/JHEP07(2020)005}{\emph{JHEP} {\bfseries 07}
  (2020) 005} [\href{https://arxiv.org/abs/1912.02188}{{\ttfamily
  1912.02188}}].

\bibitem{Nason:1995np}
P.~Nason and M.~H. Seymour, \emph{{Infrared renormalons and power suppressed
  effects in e+ e- jet events}},
  \href{https://doi.org/10.1016/0550-3213(95)00461-Z}{\emph{Nucl. Phys. B}
  {\bfseries 454} (1995) 291}
  [\href{https://arxiv.org/abs/hep-ph/9506317}{{\ttfamily hep-ph/9506317}}].

\bibitem{Korchemsky:1999kt}
G.~P. Korchemsky and G.~F. Sterman, \emph{{Power corrections to event shapes
  and factorization}},
  \href{https://doi.org/10.1016/S0550-3213(99)00308-9}{\emph{Nucl. Phys. B}
  {\bfseries 555} (1999) 335}
  [\href{https://arxiv.org/abs/hep-ph/9902341}{{\ttfamily hep-ph/9902341}}].

\bibitem{Belitsky:2001gf}
A.~V. Belitsky, G.~P. Korchemsky and G.~F. Sterman, \emph{{Energy flow in QCD
  and event shape functions}}, {\emph{Phys. Lett. B} {\bfseries 515} (2001)
  297} [\href{https://arxiv.org/abs/hep-ph/0106308}{{\ttfamily
  hep-ph/0106308}}].

\bibitem{Dokshitzer:1999py}
Y.~L. Dokshitzer, G.~Marchesini and B.~R. Webber, \emph{{Nonperturbative
  effects in the energy energy correlation}}, {\emph{JHEP} {\bfseries 07}
  (1999) 012} [\href{https://arxiv.org/abs/hep-ph/9905339}{{\ttfamily
  hep-ph/9905339}}].

\bibitem{Dokshitzer:1998pt}
Y.~L. Dokshitzer, A.~Lucenti, G.~Marchesini and G.~P. Salam, \emph{{On the
  universality of the Milan factor for 1 / Q power corrections to jet shapes}},
  \href{https://doi.org/10.1088/1126-6708/1998/05/003}{\emph{JHEP} {\bfseries
  05} (1998) 003} [\href{https://arxiv.org/abs/hep-ph/9802381}{{\ttfamily
  hep-ph/9802381}}].

\bibitem{Dokshitzer:1997iz}
Y.~L. Dokshitzer, A.~Lucenti, G.~Marchesini and G.~P. Salam,
  \emph{{Universality of 1/Q corrections to jet-shape observables rescued}},
  \href{https://doi.org/10.1016/S0550-3213(97)00650-0}{\emph{Nucl. Phys. B}
  {\bfseries 511} (1998) 396}
  [\href{https://arxiv.org/abs/hep-ph/9707532}{{\ttfamily hep-ph/9707532}}].

\bibitem{Abbate:2010xh}
R.~Abbate, M.~Fickinger, A.~H. Hoang, V.~Mateu and I.~W. Stewart, \emph{{Thrust
  at N$^3$LL with Power Corrections and a Precision Global Fit for
  alphas(mZ)}},
  \href{https://doi.org/10.1103/PhysRevD.83.074021}{\emph{Phys.Rev.} {\bfseries
  D83} (2011) 074021} [\href{https://arxiv.org/abs/1006.3080}{{\ttfamily
  1006.3080}}].

\bibitem{Lee:2006nr}
C.~Lee and G.~Sterman, \emph{Momentum flow correlations from event shapes:
  Factorized soft gluons and soft-collinear effective theory}, {\emph{Phys.
  Rev.} {\bfseries D75} (2007) 014022}
  [\href{https://arxiv.org/abs/hep-ph/0611061}{{\ttfamily hep-ph/0611061}}].

\bibitem{Salam:2001bd}
G.~P. Salam and D.~Wicke, \emph{{Hadron masses and power corrections to event
  shapes}}, {\emph{JHEP} {\bfseries 05} (2001) 061}
  [\href{https://arxiv.org/abs/hep-ph/0102343}{{\ttfamily hep-ph/0102343}}].

\bibitem{Mateu:2012nk}
V.~Mateu, I.~W. Stewart and J.~Thaler, \emph{{Power Corrections to Event Shapes
  with Mass-Dependent Operators}},
  \href{https://doi.org/10.1103/PhysRevD.87.014025}{\emph{Phys. Rev. D}
  {\bfseries 87} (2013) 014025}
  [\href{https://arxiv.org/abs/1209.3781}{{\ttfamily 1209.3781}}].

\bibitem{Hoang:2009yr}
A.~H. Hoang, A.~Jain, I.~Scimemi and I.~W. Stewart, \emph{{R-evolution:
  Improving perturbative QCD}}, {\emph{Phys. Rev. D} {\bfseries 82} (2010)
  011501} [\href{https://arxiv.org/abs/0908.3189}{{\ttfamily 0908.3189}}].

\bibitem{Bachu:2020nqn}
B.~Bachu, A.~H. Hoang, V.~Mateu, A.~Pathak and I.~W. Stewart, \emph{{Boosted
  top quarks in the peak region with NL3L resummation}},
  \href{https://doi.org/10.1103/PhysRevD.104.014026}{\emph{Phys. Rev. D}
  {\bfseries 104} (2021) 014026}
  [\href{https://arxiv.org/abs/2012.12304}{{\ttfamily 2012.12304}}].

\bibitem{OPAL:1993pnw}
{\scshape OPAL} collaboration, P.~D. Acton et~al., \emph{{A Determination of
  alpha-s (M (Z0)) at LEP using resummed QCD calculations}}, {\emph{Z. Phys. C}
  {\bfseries 59} (1993) 1}.

\bibitem{Dyson:1952tj}
F.~J. Dyson, \emph{{Divergence of perturbation theory in quantum
  electrodynamics}}, \href{https://doi.org/10.1103/PhysRev.85.631}{\emph{Phys.
  Rev.} {\bfseries 85} (1952) 631}.

\bibitem{Bender:1969si}
C.~M. Bender and T.~T. Wu, \emph{{Anharmonic oscillator}},
  \href{https://doi.org/10.1103/PhysRev.184.1231}{\emph{Phys. Rev.} {\bfseries
  184} (1969) 1231}.

\bibitem{Bender:1973rz}
C.~M. Bender and T.~T. Wu, \emph{{Anharmonic oscillator. 2: A Study of
  perturbation theory in large order}},
  \href{https://doi.org/10.1103/PhysRevD.7.1620}{\emph{Phys. Rev. D} {\bfseries
  7} (1973) 1620}.

\bibitem{Beneke:1998ui}
M.~Beneke, \emph{{Renormalons}},
  \href{https://doi.org/10.1016/S0370-1573(98)00130-6}{\emph{Phys. Rept.}
  {\bfseries 317} (1999) 1}
  [\href{https://arxiv.org/abs/hep-ph/9807443}{{\ttfamily hep-ph/9807443}}].

\bibitem{Argyres:2012ka}
P.~C. Argyres and M.~Unsal, \emph{{The semi-classical expansion and resurgence
  in gauge theories: new perturbative, instanton, bion, and renormalon
  effects}}, \href{https://doi.org/10.1007/JHEP08(2012)063}{\emph{JHEP}
  {\bfseries 08} (2012) 063} [\href{https://arxiv.org/abs/1206.1890}{{\ttfamily
  1206.1890}}].

\bibitem{Dunne:2013ada}
G.~V. Dunne and M.~\"Unsal, \emph{{Generating nonperturbative physics from
  perturbation theory}}, {\emph{Phys. Rev. D} {\bfseries 89} (2014) 041701}
  [\href{https://arxiv.org/abs/1306.4405}{{\ttfamily 1306.4405}}].

\bibitem{Hurst:1952zh}
C.~A. Hurst, \emph{{The Enumeration of Graphs in the Feynman-Dyson Technique}},
  \href{https://doi.org/10.1098/rspa.1952.0149}{\emph{Proc. Roy. Soc. Lond. A}
  {\bfseries 214} (1952) 44}.

\bibitem{Bender:1976ni}
C.~M. Bender and T.~T. Wu, \emph{{Statistical Analysis of Feynman Diagrams}},
  \href{https://doi.org/10.1103/PhysRevLett.37.117}{\emph{Phys. Rev. Lett.}
  {\bfseries 37} (1976) 117}.

\bibitem{Lipatov:1976ny}
L.~N. Lipatov, \emph{{Divergence of the Perturbation Theory Series and the
  Quasiclassical Theory}}, {\emph{Sov. Phys. JETP} {\bfseries 45} (1977) 216}.

\bibitem{Zinn-Justin:1980oco}
J.~Zinn-Justin, \emph{{Perturbation Series at Large Orders in Quantum Mechanics
  and Field Theories: Application to the Problem of Resummation}},
  \href{https://doi.org/10.1016/0370-1573(81)90016-8}{\emph{Phys. Rept.}
  {\bfseries 70} (1981) 109}.

\bibitem{Bender_Orszag_1999}
C.~M. Bender and S.~A. Orszag, \emph{Advanced Mathematical Methods for
  Scientists and Engineers I}. Springer New York, 1999,
  \href{https://doi.org/10.1007/978-1-4757-3069-2}{10.1007/978-1-4757-3069-2}.

\bibitem{Gross:1974jv}
D.~J. Gross and A.~Neveu, \emph{{Dynamical Symmetry Breaking in Asymptotically
  Free Field Theories}},
  \href{https://doi.org/10.1103/PhysRevD.10.3235}{\emph{Phys. Rev. D}
  {\bfseries 10} (1974) 3235}.

\bibitem{Lautrup:1977hs}
B.~E. Lautrup, \emph{{On High Order Estimates in QED}},
  \href{https://doi.org/10.1016/0370-2693(77)90145-9}{\emph{Phys. Lett. B}
  {\bfseries 69} (1977) 109}.

\bibitem{tHooft:1977xjm}
G.~'t~Hooft, \emph{{Can We Make Sense Out of Quantum Chromodynamics?}},
  {\emph{Subnucl. Ser.} {\bfseries 15} (1979) 943}.

\bibitem{Bogomolny:1977ty}
E.~B. Bogomolny and V.~A. Fateev, \emph{{Large Orders Calculations in the Gauge
  Theories}}, \href{https://doi.org/10.1016/0370-2693(77)90748-1}{\emph{Phys.
  Lett. B} {\bfseries 71} (1977) 93}.

\bibitem{Behtash:2018voa}
A.~Behtash, G.~V. Dunne, T.~Schaefer, T.~Sulejmanpasic and M.~\"Unsal,
  \emph{{Critical Points at Infinity, Non-Gaussian Saddles, and Bions}},
  \href{https://doi.org/10.1007/JHEP06(2018)068}{\emph{JHEP} {\bfseries 06}
  (2018) 068} [\href{https://arxiv.org/abs/1803.11533}{{\ttfamily
  1803.11533}}].

\bibitem{Dunne:2012zk}
G.~V. Dunne and M.~\"Unsal, \emph{{Continuity and Resurgence: towards a
  continuum definition of the $\mathbb{CP}$(N-1) model}},
  \href{https://doi.org/10.1103/PhysRevD.87.025015}{\emph{Phys. Rev. D}
  {\bfseries 87} (2013) 025015}
  [\href{https://arxiv.org/abs/1210.3646}{{\ttfamily 1210.3646}}].

\bibitem{Cherman:2013yfa}
A.~Cherman, D.~Dorigoni, G.~V. Dunne and M.~\"Unsal, \emph{{Resurgence in
  Quantum Field Theory: Nonperturbative Effects in the Principal Chiral
  Model}}, \href{https://doi.org/10.1103/PhysRevLett.112.021601}{\emph{Phys.
  Rev. Lett.} {\bfseries 112} (2014) 021601}
  [\href{https://arxiv.org/abs/1308.0127}{{\ttfamily 1308.0127}}].

\bibitem{Basar:2013eka}
G.~Basar, G.~V. Dunne and M.~Unsal, \emph{{Resurgence theory, ghost-instantons,
  and analytic continuation of path integrals}},
  \href{https://doi.org/10.1007/JHEP10(2013)041}{\emph{JHEP} {\bfseries 10}
  (2013) 041} [\href{https://arxiv.org/abs/1308.1108}{{\ttfamily 1308.1108}}].

\bibitem{Dunne:2014bca}
G.~V. Dunne and M.~Unsal, \emph{{Uniform WKB, Multi-instantons, and Resurgent
  Trans-Series}}, \href{https://doi.org/10.1103/PhysRevD.89.105009}{\emph{Phys.
  Rev. D} {\bfseries 89} (2014) 105009}
  [\href{https://arxiv.org/abs/1401.5202}{{\ttfamily 1401.5202}}].

\bibitem{Cherman:2014ofa}
A.~Cherman, D.~Dorigoni and M.~Unsal, \emph{{Decoding perturbation theory using
  resurgence: Stokes phenomena, new saddle points and Lefschetz thimbles}},
  \href{https://doi.org/10.1007/JHEP10(2015)056}{\emph{JHEP} {\bfseries 10}
  (2015) 056} [\href{https://arxiv.org/abs/1403.1277}{{\ttfamily 1403.1277}}].

\bibitem{Fujimori:2018kqp}
T.~Fujimori, S.~Kamata, T.~Misumi, M.~Nitta and N.~Sakai, \emph{{Bion
  non-perturbative contributions versus infrared renormalons in two-dimensional
  $\mathbb C P^{N-1}$ models}},
  \href{https://doi.org/10.1007/JHEP02(2019)190}{\emph{JHEP} {\bfseries 02}
  (2019) 190} [\href{https://arxiv.org/abs/1810.03768}{{\ttfamily
  1810.03768}}].

\bibitem{Unsal:2021cch}
M.~\"Unsal, \emph{{TQFT at work for IR-renormalons, resurgence and Lefschetz
  decomposition}},  \href{https://arxiv.org/abs/2106.14971}{{\ttfamily
  2106.14971}}.

\bibitem{Ecalle:1981:FRTa}
J.~{\'E}calle, \emph{Les fonctions r\'esurgentes I-III}. Universit\'e de
  Paris-Sud D\'epartement de Math\'ematique, Orsay, 1981.

\bibitem{boyd_devils_1999}
J.~P. Boyd, \emph{The {Devil}'s {Invention}: {Asymptotic}, {Superasymptotic}
  and {Hyperasymptotic} {Series}},
  \href{https://doi.org/10.1023/A:1006145903624}{\emph{Acta Applicandae
  Mathematica} {\bfseries 56} (1999) 1}.

\bibitem{Dorigoni:2014hea}
D.~Dorigoni, \emph{{An Introduction to Resurgence, Trans-Series and Alien
  Calculus}}, \href{https://doi.org/10.1016/j.aop.2019.167914}{\emph{Annals
  Phys.} {\bfseries 409} (2019) 167914}
  [\href{https://arxiv.org/abs/1411.3585}{{\ttfamily 1411.3585}}].

\bibitem{Aniceto:2018bis}
I.~Aniceto, G.~Basar and R.~Schiappa, \emph{{A Primer on Resurgent Transseries
  and Their Asymptotics}},
  \href{https://doi.org/10.1016/j.physrep.2019.02.003}{\emph{Phys. Rept.}
  {\bfseries 809} (2019) 1} [\href{https://arxiv.org/abs/1802.10441}{{\ttfamily
  1802.10441}}].

\bibitem{Manohar:2000dt}
A.~V. Manohar and M.~B. Wise, \emph{{Heavy quark physics}}, vol.~10. 2000.

\bibitem{Bigi:1994ng}
I.~I.~Y. Bigi, M.~A. Shifman, N.~G. Uraltsev and A.~I. Vainshtein, \emph{{The
  Pole mass of the heavy quark. Perturbation theory and beyond}}, {\emph{Phys.
  Rev. D} {\bfseries 50} (1994) 2234}
  [\href{https://arxiv.org/abs/hep-ph/9402360}{{\ttfamily hep-ph/9402360}}].

\bibitem{Beneke:1994sw}
M.~Beneke and V.~M. Braun, \emph{{Heavy quark effective theory beyond
  perturbation theory: Renormalons, the pole mass and the residual mass term}},
  {\emph{Nucl. Phys. B} {\bfseries 426} (1994) 301}
  [\href{https://arxiv.org/abs/hep-ph/9402364}{{\ttfamily hep-ph/9402364}}].

\bibitem{Bigi:1997ty}
I.~I.~Y. Bigi, M.~A. Shifman and N.~Uraltsev, \emph{{Aspects of heavy quark
  theory}}, {\emph{Ann. Rev. Nucl. Part. Sci.} {\bfseries 47} (1997) 591}
  [\href{https://arxiv.org/abs/hep-ph/9703290}{{\ttfamily hep-ph/9703290}}].

\bibitem{Pineda:2001ia}
A.~Pineda, \emph{{Determination of the bottom quark mass from the Upsilon(1S)
  system}}, {\emph{JHEP} {\bfseries 06} (2001) 022}
  [\href{https://arxiv.org/abs/hep-ph/0105008}{{\ttfamily hep-ph/0105008}}].

\bibitem{Hoang:2008yj}
A.~H. Hoang, A.~Jain, I.~Scimemi and I.~W. Stewart, \emph{{Infrared
  Renormalization Group Flow for Heavy Quark Masses}}, {\emph{Phys. Rev. Lett.}
  {\bfseries 101} (2008) 151602}
  [\href{https://arxiv.org/abs/0803.4214}{{\ttfamily 0803.4214}}].

\bibitem{Grozin:1997sa}
A.~G. Grozin and M.~Neubert, \emph{{Higher order estimates of the
  chromomagnetic moment of a heavy quark}}, {\emph{Nucl. Phys. B} {\bfseries
  508} (1997) 311} [\href{https://arxiv.org/abs/hep-ph/9707318}{{\ttfamily
  hep-ph/9707318}}].

\bibitem{Mueller:1984vh}
A.~H. Mueller, \emph{{On the Structure of Infrared Renormalons in Physical
  Processes at High-Energies}}, {\emph{Nucl. Phys. B} {\bfseries 250} (1985)
  327}.

\bibitem{Dasgupta:2003iq}
M.~Dasgupta and G.~P. Salam, \emph{{Event shapes in e+ e- annihilation and deep
  inelastic scattering}},
  \href{https://doi.org/10.1088/0954-3899/30/5/R01}{\emph{J. Phys. G}
  {\bfseries 30} (2004) R143}
  [\href{https://arxiv.org/abs/hep-ph/0312283}{{\ttfamily hep-ph/0312283}}].

\bibitem{Fleming:2008po}
S.~Fleming, A.~H. Hoang, S.~Mantry and I.~W. Stewart, \emph{{Top Jets in the
  Peak Region: Factorization Analysis with NLL Resummation}}, {\emph{Phys. Rev.
  D} {\bfseries 77} (2008) 114003}
  [\href{https://arxiv.org/abs/0711.2079}{{\ttfamily 0711.2079}}].

\bibitem{Hoang:2015gj}
A.~H. Hoang, D.~W. Kolodrubetz, V.~Mateu and I.~W. Stewart,
  \emph{{$C$-parameter distribution at N$^3$LL' including power corrections}},
  {\emph{Phys. Rev. D} {\bfseries 91} (2015) 094017}
  [\href{https://arxiv.org/abs/1411.6633}{{\ttfamily 1411.6633}}].

\bibitem{Gracia:2021nut}
N.~G. Gracia and V.~Mateu, \emph{{Toward massless and massive event shapes in
  the large-\ensuremath{\beta}$_{0}$ limit}},
  \href{https://doi.org/10.1007/JHEP07(2021)229}{\emph{JHEP} {\bfseries 07}
  (2021) 229} [\href{https://arxiv.org/abs/2104.13942}{{\ttfamily
  2104.13942}}].

\bibitem{Brodsky:1982gc}
S.~J. Brodsky, G.~P. Lepage and P.~B. Mackenzie, \emph{{On the Elimination of
  Scale Ambiguities in Perturbative Quantum Chromodynamics}},
  \href{https://doi.org/10.1103/PhysRevD.28.228}{\emph{Phys. Rev. D} {\bfseries
  28} (1983) 228}.

\bibitem{Hoang:2007vb}
A.~H. Hoang and I.~W. Stewart, \emph{{Designing gapped soft functions for jet
  production}}, {\emph{Phys. Lett. B} {\bfseries 660} (2008) 483}
  [\href{https://arxiv.org/abs/0709.3519}{{\ttfamily 0709.3519}}].

\bibitem{Kolouglu:2021lig}
M.~Kolo{\u{g}}lu, P.~Kravchuk, D.~Simmons-Duffin and A.~Zhiboedov, \emph{The
  light-ray ope and conformal colliders}, {\emph{Journal of High Energy
  Physics} {\bfseries 2021} (2021) 1}.

\bibitem{Gardi:2001ny}
E.~Gardi and J.~Rathsman, \emph{{Renormalon resummation and exponentiation of
  soft and collinear gluon radiation in the thrust distribution}},
  \href{https://doi.org/10.1016/S0550-3213(01)00284-X}{\emph{Nucl. Phys. B}
  {\bfseries 609} (2001) 123}
  [\href{https://arxiv.org/abs/hep-ph/0103217}{{\ttfamily hep-ph/0103217}}].

\bibitem{Hoang:2008fs}
A.~H. Hoang and S.~Kluth, \emph{{Hemisphere Soft Function at O($\alpha_s^2$)
  for Dijet Production in e+e- Annihilation}},
  \href{https://arxiv.org/abs/0806.3852}{{\ttfamily 0806.3852}}.

\bibitem{Bruser:2018rad}
R.~Br\"user, Z.~L. Liu and M.~Stahlhofen, \emph{{Three-Loop Quark Jet
  Function}}, {\emph{Phys. Rev. Lett.} {\bfseries 121} (2018) 072003}
  [\href{https://arxiv.org/abs/1804.09722}{{\ttfamily 1804.09722}}].

\bibitem{Hoang:2017suc}
A.~H. Hoang, A.~Jain, C.~Lepenik, V.~Mateu, M.~Preisser, I.~Scimemi et~al.,
  \emph{{The MSR mass and the $
  \mathcal{O}\left({\Lambda}_{\mathrm{QCD}}\right) $ renormalon sum rule}},
  {\emph{JHEP} {\bfseries 04} (2018) 003}
  [\href{https://arxiv.org/abs/1704.01580}{{\ttfamily 1704.01580}}].

\bibitem{Chetyrkin:1996ela}
K.~G. Chetyrkin, J.~H. Kuhn and A.~Kwiatkowski, \emph{{QCD corrections to the
  $e^{+} e^{-}$ cross-section and the $Z$ boson decay rate}},
  \href{https://doi.org/10.1016/S0370-1573(96)00012-9}{\emph{Phys. Rept.}
  {\bfseries 277} (1996) 189}
  [\href{https://arxiv.org/abs/hep-ph/9503396}{{\ttfamily hep-ph/9503396}}].

\end{thebibliography}\endgroup

\end{document}